\documentclass[twocolumn,superscriptaddress,nofootinbib,floatfix]{revtex4-1}
\usepackage{epsfig}
\usepackage{amsmath}
\usepackage{amsfonts}
\usepackage{float}
\usepackage{amssymb}
\usepackage{color}
\usepackage{pbox}
\usepackage{subfigure}
\usepackage{array,multirow,makecell}
\usepackage[colorlinks,citecolor=blue]{hyperref}
\usepackage{tabularx}
\usepackage{titlesec}
\usepackage{ulem}
\usepackage{natbib}
\usepackage{array,multirow,makecell}
\usepackage{lipsum}
\usepackage[colorlinks,citecolor=blue]{hyperref}

\def\lsim{\raise0.3ex\hbox{$\;<$\kern-0.75em\raise-1.1ex\hbox{$\sim\;$}}}
\def\gsim{\raise0.3ex\hbox{$\;>$\kern-0.75em\raise-1.1ex\hbox{$\sim\;$}}}
\def\be{\begin{equation}}
\def\ee{\end{equation}}
\def\bea{\begin{eqnarray}}
\def\eea{\end{eqnarray}}

\titleformat{\paragraph}
{\normalfont\normalsize\bfseries}{\theparagraph}{1em}{}
\titlespacing*{\paragraph}
{0pt}{3.25ex plus 1ex minus .2ex}{1.5ex plus .2ex}
\begin{document}

\begin{flushright}
HRI-RECAPP-2021-001
\end{flushright} 

\title{Scalar Multiplet Dark Matter in a Fast Expanding Universe: resurrection  of the {\it desert} region}

\author{Basabendu Barman}
\email[E-mail: ]{bb1988@iitg.ac.in}
\affiliation{Department of Physics, IIT Guwahati, Guwahati-781039, India}

\author{Purusottam Ghosh}
\email[E-mail: ]{purusottamghosh@hri.res.in }
\affiliation{Regional Centre for Accelerator-based Particle Physics, Harish-Chandra Research Institute, HBNI, Chhatnag Road, Jhunsi, Allahabad-211019, India}

\author{Farinaldo S. Queiroz}
\email[E-mail: ]{farinaldo.queiroz@iip.ufrn.br}
\affiliation{International Institute of Physics, Universidade Federal do Rio Grande do Norte, Campus Universit\'ario, Lagoa Nova, Natal-RN 59078-970, Brazil\\
Departamento de F\'isica, Universidade Federal do Rio Grande do Norte, 59078-970, Natal, RN, Brasil}

\author{Abhijit Kumar Saha}
\email[E-mail: ]{psaks2484@iacs.res.in}

\affiliation{School of Physical Sciences, Indian Association for the Cultivation of Science, 2A $\&$ 2B Raja S.C. Mullick Road, Kolkata 700 032, India}

%

\begin{abstract}
{\bf Abstract:} We examine the impact of a faster expanding Universe on the phenomenology of scalar  dark matter (DM) associated with $SU(2)_L$ multiplets. Earlier works with radiation dominated Universe have reported the presence of {\it {\it desert}} region for both inert $SU(2)_L$ doublet and triplet DM candidates where the DM is under abundant. We find that the existence of a faster expanding component before BBN can revive a substantial part of the {\it desert} parameter space consistent with relic density requirements and other direct and indirect search bounds. We also review the possible collider search prospects of the newly obtained parameter space and predict that such region might be probed at the future colliders with improved sensitivity via a disappearing/stable charged track.

\end{abstract}
\maketitle
\section{Introduction}
\label{sec:intro}








Production of dark matter (DM) in scenarios with a non-standard history has gained growing interest in recent  times~\cite{Chung:1998rq,Okada:2004nc,Okada:2007na,Okada:2009xe,Allahverdi:2018aux,Baules:2019zwk,Waldstein:2016blt,Arias:2019uol,Cosme:2020nac,Aparicio:2016qqb,Han:2019vxi,Drees:2017iod,Arcadi:2020aot,Cosme:2020mck,Bernal:2019mhf,Allahverdi:2019jsc,Maldonado:2019qmp,Drees:2018dsj,Bernal:2018ins,Visinelli:2017qga,Arbey:2018uho,Berger:2020maa,McDonald:1989jd,Poulin:2019omz,Hardy:2018bph,Redmond:2017tja,DEramo:2017gpl,DEramo:2017ecx,Bernal:2018kcw,Chanda:2019xyl,Gelmini:2019esj,Biswas:2018iny,Fernandez:2018tfa,Betancur:2018xtj,Mahanta:2019sfo,Allahverdi:2020bys}. Since, the cosmological history of the early Universe prior to Big Bang Nucleosynthesis (BBN) is vastly dark, the possibility of presence of a non standard era in the early Universe is open. In fact, there are no fundamental reasons to assume that the Universe was radiation-dominated (RD) in the pre-BBN regime at $t\sim 1~\rm sec$. The history of the Universe can be modelled, in general, by the presence of a fluid with arbitrary equation of state parameter, which is zero for matter domination. If the equation of state parameter of a fluid turns out to be larger than the value for radiation, then the fluid acts as a fast expanding component.

Study of DM phenomenology in the presence of a modified cosmological epoch has been performed widely and it shows several significant observational consequences~\cite{Artymowski:2016tme,Hardy:2018bph,Redmond:2017tja}. In~\cite{DEramo:2017gpl}, a model independent analysis of DM phenomenology in a fast expanding Universe is worked out. It has been observed that if DM freezes out during the fast expansion of the Universe, the required interaction strength increases than the one in the standard scenario in order to satisfy the relic bound by PLANCK experiment. At some stage during the evolution of the Universe, at least before the BBN, the domination of fast expanding component has to end such that the standard RD Universe takes over. A similar phenomenological study with freeze-in production of DM in a fast expanding Universe has been explored in~\cite{DEramo:2017ecx}. With the emergence of this proposal, further efforts  have been put forward to cultivate the DM phenomenology considering such non-standard scenario in different well established beyond standard model frameworks. For example, phenomenology of a real gauge singlet scalar DM in non standard cosmologies can be found in~\cite{Bernal:2018kcw}. Well motivated anatomy on the revival of $Z$-boson and Higgs mediated DM model with alternative cosmology (late matter decay) are presented in~\cite{Chanda:2019xyl,Hardy:2018bph}. In~\cite{Arcadi:2020aot,Biswas:2018iny,Fernandez:2018tfa} the possibility of sterile neutrinos as dark matter candidates with modified cosmology have been discussed. Such sterile neutrinos can provide a sensitive probe of the pre-BBN epoch as pointed out in \cite{Gelmini:2019esj}. In~\cite{Betancur:2018xtj} the case for fermion DM originating from different order of multiplets is studied. 


Motivated from these, in the present work, we aim to resurrect the so called {\it desert} region in the parameter space of the $SU(2)_L$ inert doublet (IDM) and triplet dark matter (ITDM) models \cite{Cirelli:2005uq,Hambye:2009pw} by considering the presence of a faster expanding component (kinaton or faster than kinaton) in the early Universe. In the context of single component\footnote{In multi-component DM framework,  individual DM candidates can be under abundant, and the {\it desert} region is thus not an issue there. Such frameworks involving multi-component DM are proposed in \cite{Bhattacharya:2019fgs, Chakrabarty:2021kmr,DuttaBanik:2020jrj}. } IDM dark matter it is well known~\cite{LopezHonorez:2010tb,Borah:2017dfn} that the intermediate DM mass regime $80\lesssim m_\text{DM}\lesssim 525~\rm GeV$ suffers from under abundance issue. It occurs due to large interaction rate of the DM (mediated by $SU(2)_L$ gauge bosons) with the SM particles   resulting into late freeze out and subsequently less abundance. This particular mass window for IDM is thus referred as the {\it desert} in the relic density allowed parameter space for the DM. On the other hand, for single component DM that stems from an inert scalar triplet, right relic density is achieved at a very large DM mass $\gtrsim 2~\rm TeV$ under standard freeze-out assumptions. This happens due to small radiative mass splitting between the charged and neutral component of the scalar triplet is $\sim 166~\rm MeV$ which leads to huge co-annihilation resulting in DM under abundance. Several prescriptions have been put forward for the revival of the IDM {\it desert}. These ideas basically revolve around extending the SM particle content~\cite{Borah:2017dfn,Chakraborti:2018aae}. The case for scotogenic DM model in a modified cosmological scenario has been discussed earlier in ~\cite{Mahanta:2019sfo}. Although authors of~\cite{Mahanta:2019sfo} have briefly remarked on the impact of non-standard Universe in DM relic abundance, their work is more focused on addressing neutrino mass and leptogenesis. Thus, a detailed investigation of DM phenomenology and the impact of  direct, indirect and collider searches on the DM parameter space is highly recommended. 


In the first part of the work our attempt is to make an exact prediction on the allowed parameter space of the usual IDM scenario in the presence of a fast expanding Universe. We also elucidate in detail the effect of fast expansion on the subsequent collider signature of the model. We first obtain the parameter space for the IDM dark matter that satisfies the relic abundance criteria by varying the relevant parameters that control expansion of the Universe. We find, a significant part of the relic allowed parameter space further gets disfavored upon imposing the direct and indirect search constraints  {together with the requirement of DM thermalization}, which, in turn, directly restricts the amount of fast expansion. Since the mass difference of the DM with other neutral and charged eigenstates are found to be small, the collider search of the allowed parameter space is limited and can be probed with the identification of the charged track signal of a long-lived charged scalar. We anticipate that the improved sensitivity of CMS/ATLAS search~\cite{CMS:2014gxa,Khachatryan:2015lla,Sirunyan:2018ldc} can be used as an useful tool to test the early Universe history before BBN. In the later part we extend our analysis for a $SU(2)_L$ triplet DM model with zero hypercharge. Similar to the IDM case, existence of a {\it desert} region for triplet DM is mentioned in earlier works~\cite{FileviezPerez:2008bj,Araki:2011hm,Chao:2018xwz,Jangid:2020qgo,Fiaschi:2018rky,Betancur:2017dhy,Lu:2016dbc,Lu:2016ucn,Bahrami:2015mwa}. We use the same methodology of faster-than-usual expansion to revive part of the {\it desert} confronting all possible experimental bounds (including direct and indirect searches) which has not been done earlier to the best of our knowledge.  

The paper is organised as follows: in Sec.~\ref{sec:nsc} we briefly sketch the nonstandard cosmological framework that arises due to fast expansion; the phenomenology for inert doublet DM in the light of fast expansion is elaborated in Sec.~\ref{sec:dm-fast-exp} where we have discussed the modification in the Boltzmann equation due to modified Hubble rate in subsection ~\ref{sec:dmpheno}; subsection ~\ref{sec:dm-yld} illustrates how the DM yield gets modified once fast expansion is invoked; a detailed parameter space scan showing the constraints from DM relic abundance, direct and indirect searches are discussed in subsection ~\ref{sec:dm-param-scan}; possibile collider signature for the revived parameter space is discussed in subsection ~\ref{sec:collider}; the fate of scalar triplet DM in a fast expanding Universe is illustrated in Sec.~\ref{sec:itdm} and finally in Sec.~\ref{sec:concl} we conclude by summarizing our findings.
 
\section{Nonstandard scenarios of the Universe}\label{sec:nsc}

Here we briefly present the recipe to analyze the early Universe by considering both standard and non standard scenarios. The expansion rate of the Universe measured by the Hubble parameter $\mathcal{H}$ which is connected to the total energy density of the Universe through standard Friedmann equation. In the standard case, it is assumed that the Universe was radiation dominated starting from the reheating era upto BBN. Here we assume somewhat a different possibility that the Universe before BBN were occupied by different species namely radiation and $\eta$, with  energy densities $\rho_{\rm rad}$ and $\rho_\eta$ respectively. The equation of state for a particular component is given by:

\begin{align}
 p=\omega\rho, 
\end{align}

\noindent where $p$ stands for the pressure of that component. For radiation, $\omega_R=\frac{1}{3}$, while for $\eta$, $\omega_\eta$ could be different. The $\omega_\eta=0$ case is familiar as early matter domination and $\omega_\eta=1$ is dubbed as fast expanding Universe. However irrespective of the nature of $\eta$, the energy component $\rho_\eta$ must be subdominant compared to $\rho_R$ before BBN takes place. This poses a strong lower bound on the temperature of the Universe $T\gtrsim (15.4)^{1/n}$ MeV before the onset of BBN (see Appendix.~\ref{sec:bbn}). Considering the presence of a new species ($\eta$) along with the radiation field, the total energy budget of the Universe is $\rho = \rho_{\text{rad}} + \rho_\eta$. For standard cosmology, the $\eta$ field would be absent, and we simply write $\rho=\rho_{\text{rad}}$. One can always express the energy density of the radiation component which is given by as function of temperature,

\begin{equation}
 \rho_{\text{rad}} (T) = \frac{\pi^2}{30} g_*(T)T^4,
\end{equation}

\noindent where $g_* (T)$ stands for the effective number of relativistic degrees of freedom at temperature $T$. In the limit of entropy conservation per comoving volume \textit{i.e.,} $s\,a^3=$ const., one can define $\rho_{\text{rad}}(t)\propto a(t)^{-4} $. Now, in case of a faster expansion of the Universe the energy density of $\eta$ field is anticipated to be red-shifted more rapidly than the radiation. Accordingly, one can obtain $\rho_\eta\propto a(t)^{-(4+n)}$ with $n>0$.

The entropy density of the Universe is parameterized as $s(T) = \frac{2\pi^2}{45}\,g_{*s}(T)\,T^3$ where, $g_{*s}$ is the effective relativistic degrees of freedom that contribute to the entropy density. Utilizing the energy conservation principle, a general form of $\rho_\eta$ can be constructed as:

\begin{equation}
 \rho_\eta(T) =  \rho_\eta(T_R)\,\left(\frac{g_{*s}(T)}{g_{*s}(T_R)}\right)^{(4+n)/3}\left(\frac{T}{T_R}\right)^{(4+n)}.
\end{equation}

\noindent The temperature $T_R$ is an unknown parameter ($ >T_{\text{BBN}}$) and can be safely assumed as the point of equality of two respective energy densities: $\rho_\eta(T_R)=\rho_{\text{rad}}(T_R)$. Using this criteria, it is simple to specify the total energy density at any temperature ($T>T_R$) as~\cite{DEramo:2017gpl}

\begin{align}\label{totalrho}
 \rho(T) &= \rho_{rad}(T)+\rho_{\eta}(T)\\
 &=\rho_{rad}(T)\left[1+\frac{g_* (T_R)}{g_* (T)}\left(\frac{g_{*s}(T)}{g_{*s}(T_R)}\right)^{(4+n)/3}\left(\frac{T}{T_R}\right)^n\right]\label{eq:totED}
\end{align}

\noindent From the above equation, it is evident that the energy density of the Universe at any arbitrary temperature ($T>T_R$), is dominated by $\eta$ component. Now, the standard Friedmann equation connecting the Hubble parameter with the energy density of the Universe is given by:

\bea  
\mathcal{H}^2 = \frac{\rho}{3M_{\text{Pl}}^2},
\eea

\noindent with $M_{\text{Pl}}= 2.4 \times 10^{18}$ GeV being the reduced Planck mass. At temperature higher than $T_R$ with the condition $g_*(T) = \bar g_*$ which can be considered to be some constant, the Hubble rate can approximately be recasted into the following form~\cite{DEramo:2017gpl}

\begin{align}\label{eq:mod-hubl}
 \mathcal{H}(T) &\approx \frac{\pi\bar g_*^{1/2}}{3\sqrt{10}} \frac{T^2}{M_{\text{Pl}}}\left(\frac{T}{T_R}\right)^{n/2}, ~~~~ ({\rm with ~~}T \gg T_R),\\ \nonumber
 &=\mathcal{H}_R(T)\left(\frac{T}{T_R}\right)^{n/2}, 
\end{align}

\noindent where $\mathcal{H}_R(T)\simeq 0.33~\bar{g}_*^{1/2}\frac{T^2}{M_{\rm Pl}}$, the Hubble rate for radiation dominated Universe. In case of SM, $\bar g_*$ can be identified with the total SM degrees of freedom $g_*\text{(SM)} = 106.75$. It is important to note from Eq~\eqref{eq:mod-hubl} that the expansion rate is larger than what it is supposed to be in the standard cosmological background provided $T>T_R$ and $n>0$. Hence it can be stated that if the DM freezes out during $\eta$ domination, the situation will alter significantly with respect to the one in the standard cosmology. Finally, it is worth noting that $T_R$ can not be too small such that it alters the standard BBN. For certain value of $n$, BBN constraints provide a lower limit on $T_R$ which we report in  Appendix.~\ref{sec:bbn}:

\bea
T_R \gtrsim \left(15.4\right)^{1/n}~\text{MeV}.
\label{eq:tr-bbn}
\eea

To this end, we have assumed the prescription for DM freeze-out in a fast expanding Universe in a model-agnostic way. Before moving on to the next section we would like to provide few examples where it is possible to have some physical realization of the new species $\eta$. We consider $\eta$ to be a real scalar field minimally coupled to gravity. In that case a specific form for $\omega(=p/\rho)$ can be written as
\begin{align}
    \omega=\frac{\frac{1}{2}\left(\frac{d\eta}{dt}\right)^2-V(\eta)}{\frac{1}{2}\left(\frac{d\eta}{dt}\right)^2+V(\eta)}
\end{align}
The energy density of $\eta$ redshifts as as~\cite{DEramo:2017ecx}

\bea
\rho_\eta\propto a^{-3\left(1+\omega\right)},
\eea
which can be converted to $\rho_\eta\propto a^{-4+n}$ with $\omega=\frac{1}{3}(n+1)$. For a positive scalar potential, two possible extreme limits are $\left(\frac{d\eta}{dt}\right)^2\ll V(\eta)$ or the $\left(\frac{d\eta}{dt}\right)^2\gg V(\eta)$. These correspond to $\omega\in\left(-1,+1\right)$ leading to $n\in\left(-4,+2\right)$.
The $n=2$ case is realised for a Universe dominated by kinaton which
can be identified with a quintessence fluid \cite{Caldwell:1997ii,PhysRevD.37.3406}. For theories with $n>2$ one has to consider scenarios faster than quintessence with negative potential. Example of such theories can be found in~\cite{Khoury:2001wf,Buchbinder:2007ad} where one assumes the presence of a pre big bang ``ekpyrotic" phase. The key ingredient of ekpyrosis is same as that of inflation, namely a scalar field rolling down some self-interaction potential. However, the crucial difference being, while inflation requires a flat and positive potential, its ekpyrotic counterpart is steep and negative. Note that, in this work we consider the kination or faster than kination scenario with $n\geq 2$.



\section{Scalar Multiplet Dark Matter in a fast expanding Universe}\label{sec:dm-fast-exp}

In this section we perform the phenomenological analysis of DM belonging to different representation of scalar multiplets when the Hubble parameter is modified under the assumption of faster-than-usual expansion in the pre-BBN era. Our analysis, as mentioned in the introduction, addresses two well-motivated DM scenario: 

\begin{itemize}
 \item The inert doublet model (IDM) where the second Higgs doublet carries a non-zero hypercharge and the DM emerges either as the CP-even or as the CP-odd component of the second Higgs.
 \item A hypercharge-less $(Y=0)$ inert triplet scalar under $SU(2)_L$ where the neutral component of the scalar triplet can be a viable DM candidate. We shall call this as the inert triplet dark matter (ITDM).
\end{itemize}

In either cases one has to impose a discrete symmetry to ensure the stability of the DM. The DM phenomenology for both of these models have been studied in great detail in the background of a standard radiation-dominated Universe. From this analyses it has been found that for the case of IDM the DM mass range $m_W(\sim 80)\lesssim m_\text{DM}\lesssim 525~\rm GeV$ is under abundant, while for ITDM below 1.9 TeV is under abundant. Here we would like to mention that other possibility of having a scalar triplet DM is to consider a $Y=2$ triplet, however for such a non-zero hypercharge multiplet, $Z$-mediated direct detection bound becomes severe making most of the DM parameter space forbidden simply from spin-independent direct detection bound~\cite{Araki:2011hm,Kanemura:2012rj,DuttaBanik:2020jrj}. Therefore, we shall focus only on $Y=0$ triplet. Our goal is, as emphasized earlier, to see how much of the parameter space ruled out by the standard cosmological background can be revived under the assumption of fast expansion without extending the particle spectrum for each of these models further. In the following sections we shall furnish the details of the models and explicitly demonstrate how the non-standard cosmological scenario drastically alters the standard picture. 

\subsection{The inert doublet model}
\label{sec:idm}

Here we would like to briefly summarize the inert doublet model (IDM) framework. The IDM consists of an extra scalar that transforms as a doublet under the SM gauge symmetry. An additional $Z_2$ symmetry is also imposed under which all the SM fields are even while the inert doublet transforms non-trivially. This discrete symmetry remains unbroken since it is assumed that the extra scalar does not acquire a vacuum expectation value (VEV). With this minimal particle content, the scalar potential takes the form~\cite{LopezHonorez:2006gr,LopezHonorez:2010tb,Arhrib:2013ela,Queiroz:2015utg,Belyaev:2016lok,Alves:2016bib,Borah:2017dfn,Barman:2018jhz}

\begin{align}
 V(H,\Phi)=&-\mu_H^2 |H|^2+\lambda_H|H|^4+\mu_\Phi^2(\Phi^\dagger\Phi)+\lambda_\Phi(\Phi^\dagger\Phi)^2\nonumber\\
 &+\lambda_1(H^\dagger H)(\Phi^\dagger\Phi)
 +\lambda_2(H^\dagger \Phi)(\Phi^\dagger H)\nonumber\\ &+\frac{\lambda_3}{2}\left[(H^\dagger \Phi)^2+h.c.\right].
\end{align}

After electroweak symmetry breaking (EWSB) the SM-like Higgs doublet acquires non-zero vacuum expectation value. Considering the unitary gauge, the two scalar doublets can be expressed as,

\bea\begin{aligned}
& H=\begin{pmatrix}
0 \\ \frac{h+v}{\sqrt{2}},
\end{pmatrix},~~\Phi=\begin{pmatrix}
H^\pm \\ \frac{H^0+i A^0}{\sqrt{2}},~~
\end{pmatrix}     
    \end{aligned},
\eea

\noindent where $v=246~\rm GeV$ is the SM Higgs VEV. After minimizing the potential along different field directions, one can obtain the following relations between the physical masses and the associated couplings

\bea\begin{aligned}
&  \mu_H^2=\frac{m_h^2}{2},~\mu_\Phi^2~=~m_{H^0}^2-\lambda_L v^2,~\lambda_3=\frac{1}{v^2}(m_{H^0}^2-m_{A^0}^2),\\& \lambda_2~=~\frac{1}{v^2}(m_{H^0}^2+m_{A^0}^2-2 m_{H^\pm}^2),\\& \lambda_1=2\lambda_L-\frac{2}{v^2}(m_{H^0}^2-m_{H^\pm}^2)  
    \end{aligned}\label{eq:mass-coupling}
\eea


\noindent where $\lambda_L=\frac{1}{2}(\lambda_1+\lambda_2+\lambda_3)$ and $m_h,m_{H^0},m_{A^0}$ are the mass eigenvalues of SM-like neutral scalar found at the LHC $(m_h=125.09~\text{ GeV})$, heavier or lighter additional CP-even neutral scalar and the CP-odd neutral scalar respectively. The $m_{H^\pm}$ denotes the mass of charged scalar eigenstate(s). In our case, we consider $H^0$ as the DM candidate with mass $m_{H^0}$ which automatically implies $m_{H^0}<m_{{A^0},H^\pm}$. We also assume 

\begin{align}
\Delta M=m_{A^0}-m_{H^0}=m_{H^\pm}-m_{H^0}.
\end{align}

\noindent to reduce the number of free parameters\footnote{Choosing $m_{A^0}\neq m_{H^\pm}$ does not alter our conclusions.}. Now, the masses and couplings are subject to a number of theoretical and experimental constraints. Below we briefly mention them.

\noindent $\bullet$ \textbf{Vacuum Stability:}
 Stability of the 2HDM potential is ensured by the following conditions~\cite{PhysRevD.18.2574,Ivanov:2006yq},
 \begin{gather}
   \begin{gathered}
     \lambda_H,\,\lambda_\Phi > 0\, ;
     \lambda_1 + 2\sqrt{\lambda_H\lambda_\Phi} > 0\, ;
     \\
     \lambda_1 + \lambda_2 - |\lambda_3| + 2\sqrt{\lambda_H\lambda_\Phi} > 0\, .
   \end{gathered}
   \label{4}
 \end{gather}
 
\noindent These conditions are to ensure that the scalar potential is bounded from below. 


\vspace{0.1cm}\noindent $\bullet$ \textbf{Perturbativity:}

 Tree-level unitarity imposes bounds on the size of the quartic couplings $\lambda_i$ or various combinations of them~\cite{LopezHonorez:2006gr}. On top of that, the theory remains perturbative at any given scale if naively

\bea
\left|\lambda_i\right|\lesssim 4\pi,~i=1,2,3,H,\Phi.
\eea
In view of the unitarity bound, we shall keep the magnitudes of all the relevant couplings below order of unity.

\vspace{0.1cm}\noindent $\bullet$ \textbf{Oblique parameters:}

The splitting between the heavy scalar masses is constrained by the oblique electroweak $T$-parameter~\cite{PhysRevD.46.381} whose expression in the alignment limit is given by~\cite{Barbieri:2006dq}:

\bea\begin{aligned}
& \Delta T = \frac{g_2^2}{64\pi^2 m_W^2}\Big\{\zeta\left(m_{H^\pm}^2,m_A^2\right)+\zeta\left(m_{H^\pm}^2,m_H^2\right)\\&-\zeta\left(m_A^2,m_H^2\right)\Big\},     
    \end{aligned}
\eea

with,

\bea
\zeta\left(x,y\right) = \begin{cases}                       
\frac{x+y}{2}-\frac{xy}{x-y}\ln\left(\frac{x}{y}\right), & \text{if $x\neq y$}.\\
0, & \text{if $x=y$}.\\
\end{cases}
\label{eq:t-param}
\eea

\noindent The contribution to $S$ parameter is always small~\cite{Barbieri:2006dq}, and can safely be neglected. We thus concentrate on the $T$-parameter only which is bounded by the global electroweak fit results~\cite{PhysRevD.98.030001} as

\bea
\Delta T = 0.07\pm 0.12.
\eea

It can be understood from Eq.(\ref{eq:t-param}) that the constraints on the oblique parameter typically prohibit large mass splittings among inert states. However we shall see that to satisfy the other DM related constraints, in general, relatively small mass splittings are required and hence the model easily bypasses the bounds arising from electroweak parameters.

\vspace{0.1cm}\noindent $\bullet$ \textbf{Collider bounds:}

In order to remain in compliance with the $Z$ decay width measured from LEP-II~\cite{Abbiendi:2013hk,Arbey:2017gmh} the new scalars should obey the inequality $m_Z<m_{H^0}+m_{A^0}$. The LEP experiments have performed direct searches for charged Higgs. A combination of LEP data from searches in the $\tau\nu$ and $cs$ final states demand $m_{H^\pm}\gsim 80~\rm GeV$ under the assumption that the decay $H^\pm\to W^\pm h$ is absent~\cite{Abbiendi:2013hk,Arbey:2017gmh}. As discussed in~\cite{Belanger:2015kga} Run-I of the LHC provides relevant constraints on the IDM model that significantly extend previous limits from LEP. Run-1 of ATLAS dilepton searches exclude, at 95\% CL, inert scalar masses up to about 35 GeV for pseudoscalar masses around 100 GeV, with the limits becoming stronger for larger $m_{A^0}$~\cite{Belanger:2015kga}. Also, for $m_{H^0}<m_h/2$ the SM-like CP even Higgs can decays invisibly to a pair of inert DM which is also constrained from the invisible Higgs decay width measurement at the LHC~\cite{PhysRevD.98.030001}. 

\subsubsection{IDM dark matter in the light of fast expansion}
\label{sec:dmpheno}

As stated earlier, we refer the intermediate DM mass range: $m_W\lesssim m_{H^0}\lesssim 525~\text{GeV}$ as the IDM {\it {\it desert}} where the observed relic abundance of the DM can not be generated as the DM annihilation cross section is more than what is required to produce correct abundance through the vanilla freeze-out mechanism. The inert doublet DM can (co-)annihilate to SM states through both Higgs and $Z,W^{\pm}$ mediated processes. The dominant contribution to the DM abundance generally comes from the DM pair annihilation to gauge boson final states irrespective of the choice of $\Delta M$. Although co-annihilation of DM  with its charged counterpart $H^\pm$ turns out to be important for small $\Delta M\sim 1~\rm GeV$, it provides sub-dominant contribution to the relic abundance as we have checked. Due to large annihilation rates (involving gauge interactions), the DM is under-abundant within this mass range. Without extending the model further or resorting to other DM production mechanisms, our aim is to revive the {\it desert} region with the help of non-standard cosmology. 

\begin{figure*}[htb!]
  \centering
  \subfigure[]{\includegraphics[scale=0.35]{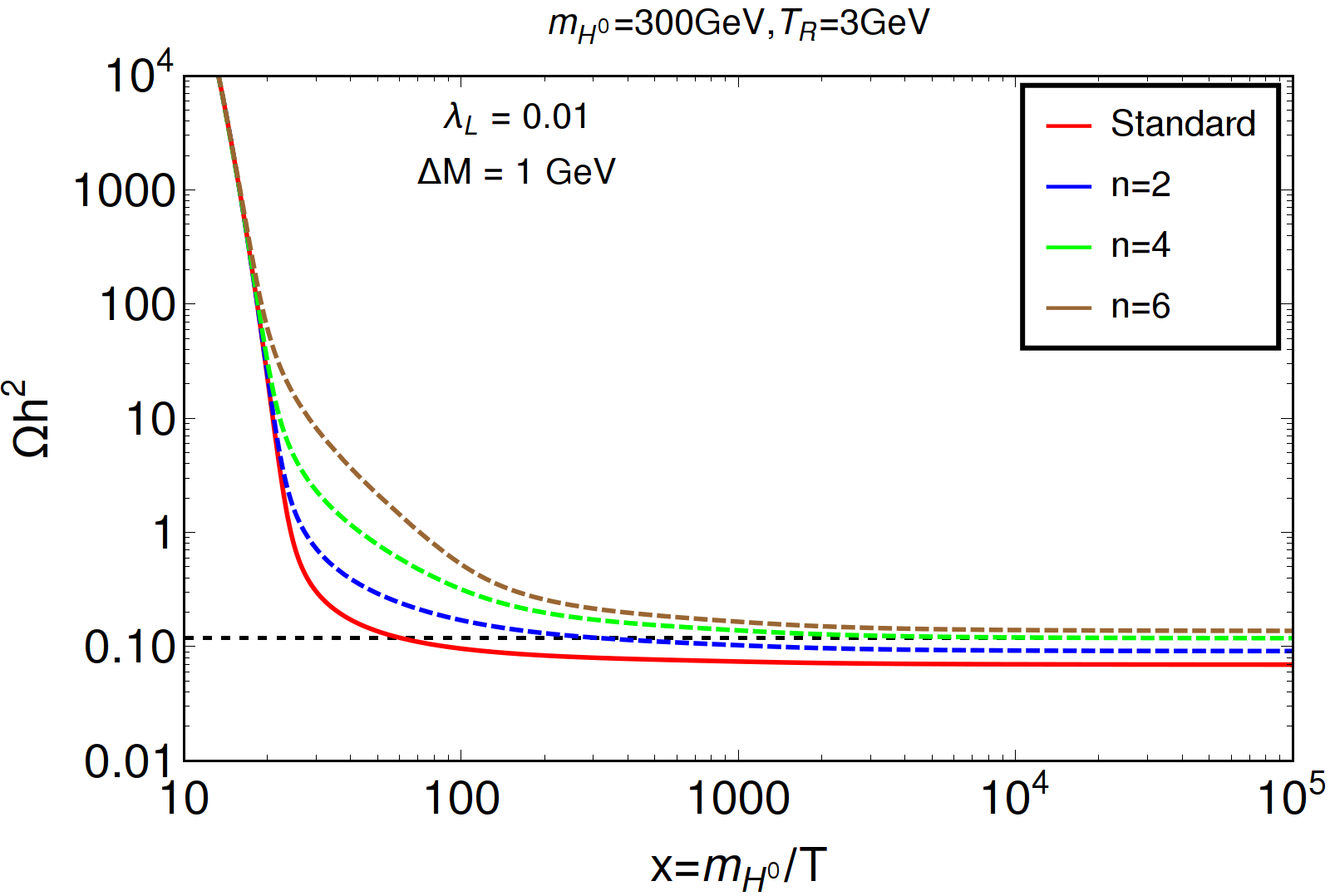}}\quad
  \subfigure[]{\includegraphics[scale=0.35]{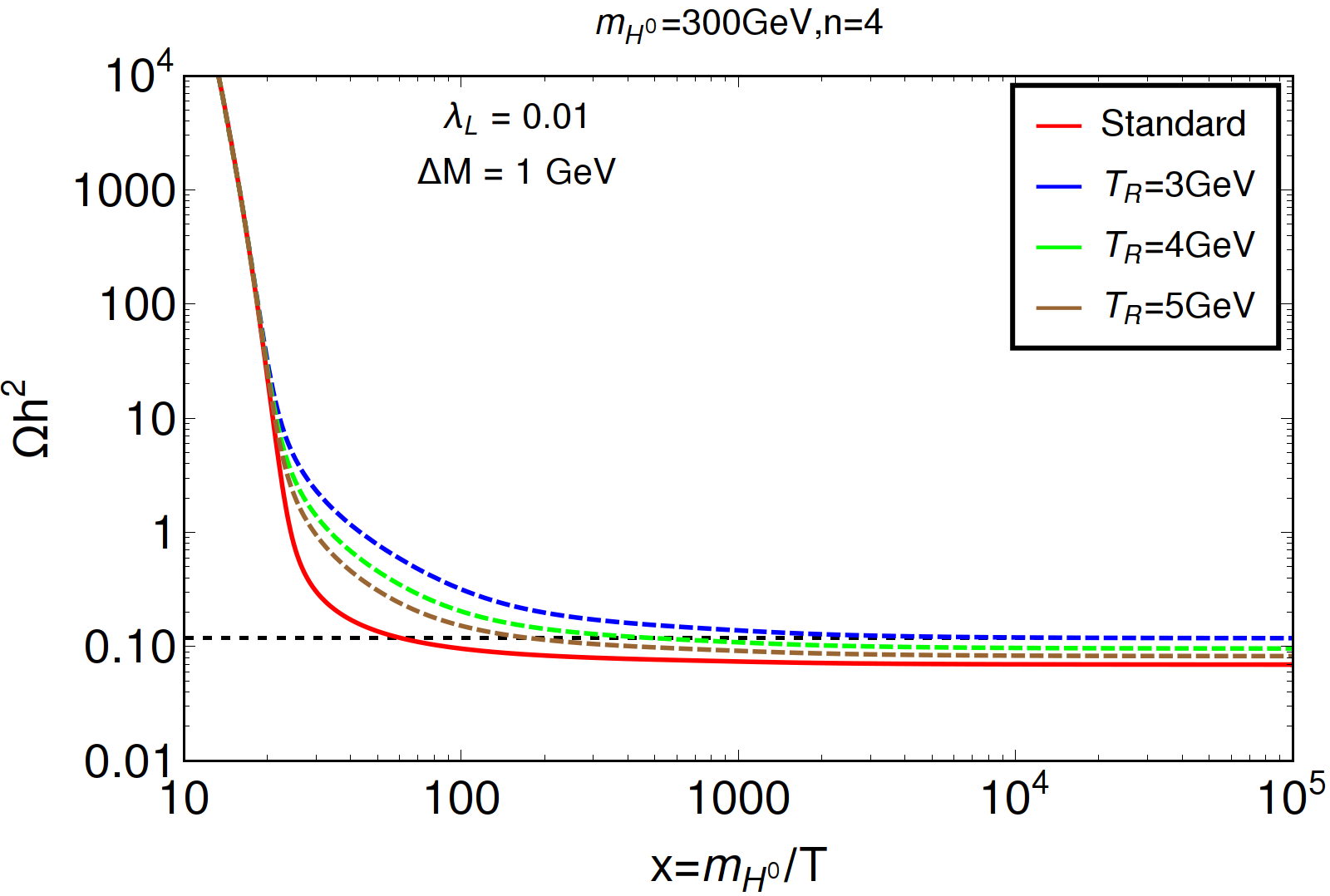}}\quad
  \caption{(a): Evolution of DM relic abundance as function of $x=m_{H^0}/T$ for RD dominated Universe (red) and in the presence of fast expansion for different values of $n(>0)$. The analysis is for a fixed $T_R=3$ GeV, $\Delta M=1~\rm GeV$ and $m_{H^0}=300~\rm GeV$ with different choices of $n=\{2,4,6\}$ shown in blue, green and brown respectively. (b): The DM relic density as a function of $x$ is plotted for a fixed $n=4$ for different choices of $T_R=\{3,4,5\}~\rm GeV$ shown via the blue, green and brown curves respectively. In both the plots the red solid curve corresponds to usual RD Universe with $n=0$ and the thick dashed straight line indicates the central value of the observed DM relic abundance.}\label{fig:dm-relic}
\end{figure*}

\begin{figure*}[htb!]
$$
 \includegraphics[height=6.1cm,width=8cm]{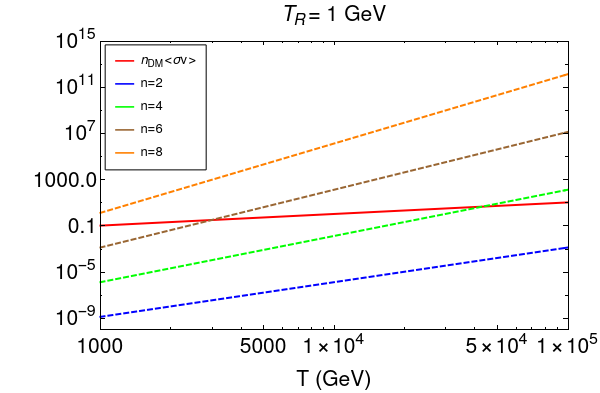}~~
 \includegraphics[height=6.2cm,width=8.3cm]{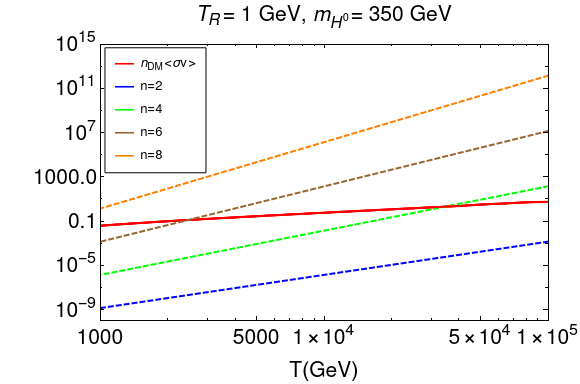}
 $$
\caption{The modified Hubble rates (dashed lines) are plotted as function of temperature for different values of $n$. The red solid line indicates the DM interaction rate $\Gamma_\text{int}$ (see text for details) as a function of temperature $T$ of the Universe. The figure in the left panel is obtained using Eq.~\eqref{eq:non-therm}, while for right panel we obtain the thermally averaged cross-section numerically to determine the DM interaction cross section.}\label{fig:Oeq-Approx}  
\end{figure*}


\begin{figure*}[htb!]
  \centering
  \subfigure[]{\includegraphics[scale=0.28]{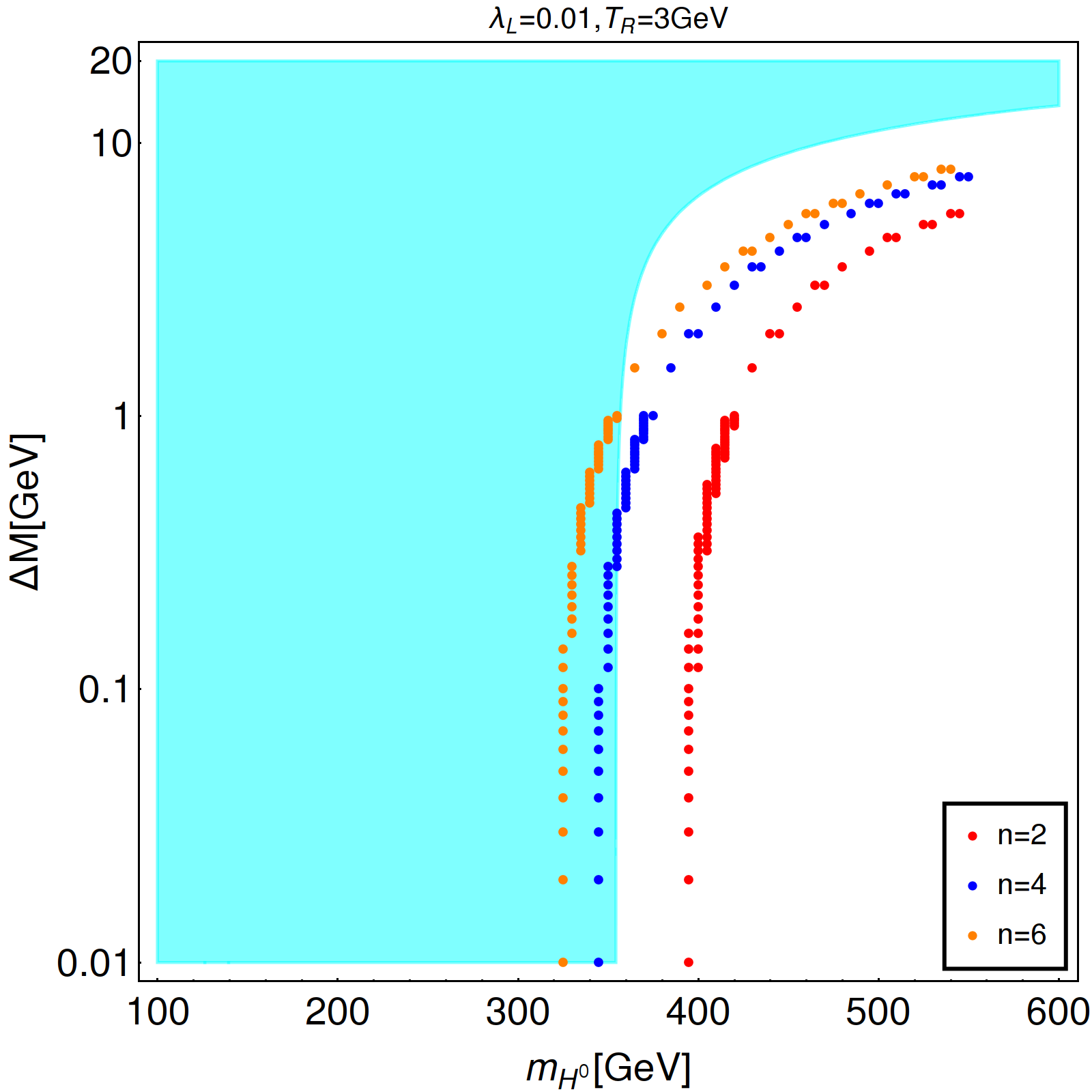}}\quad
  \subfigure[]{\includegraphics[scale=0.28]{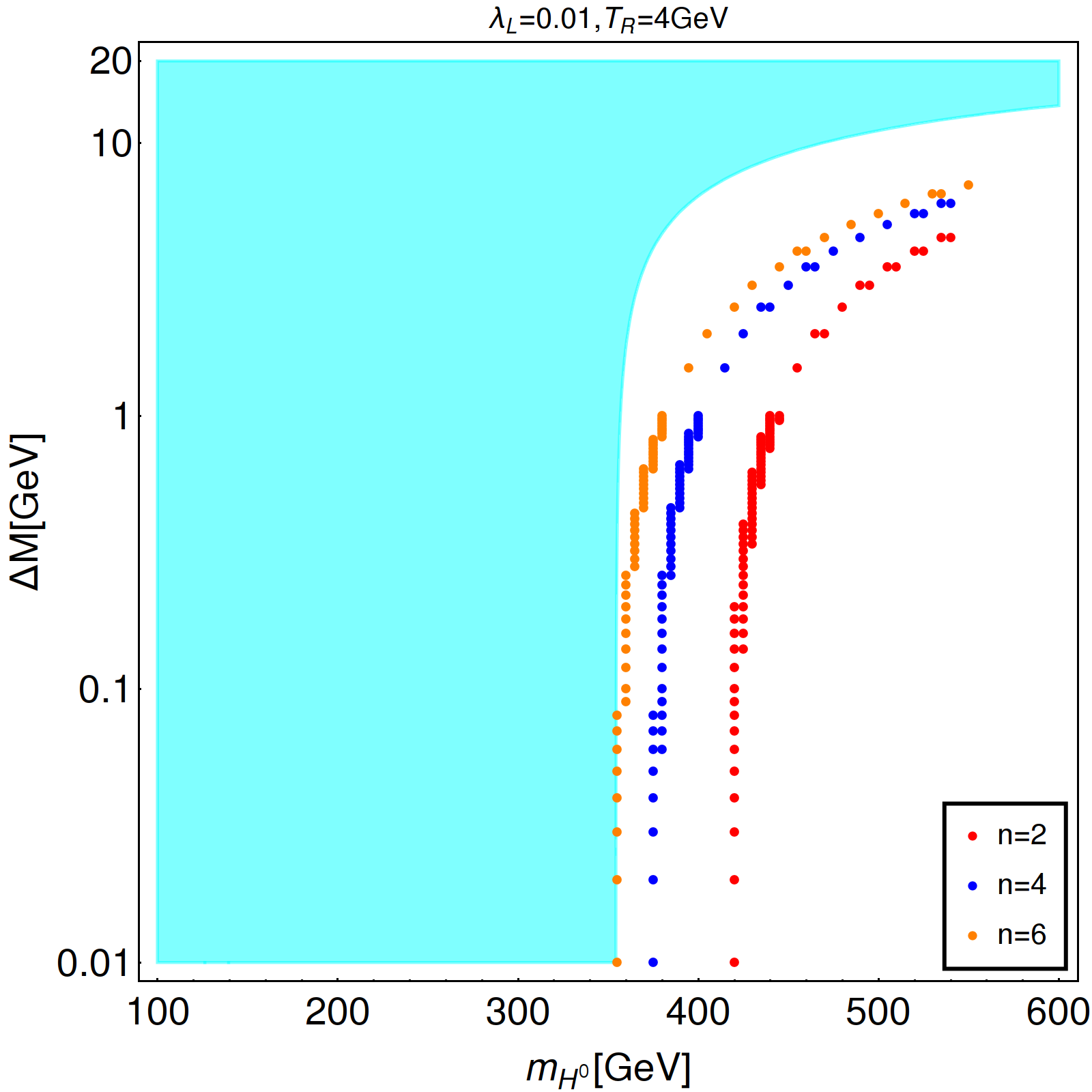}}\quad
  \subfigure[]{\includegraphics[scale=0.28]{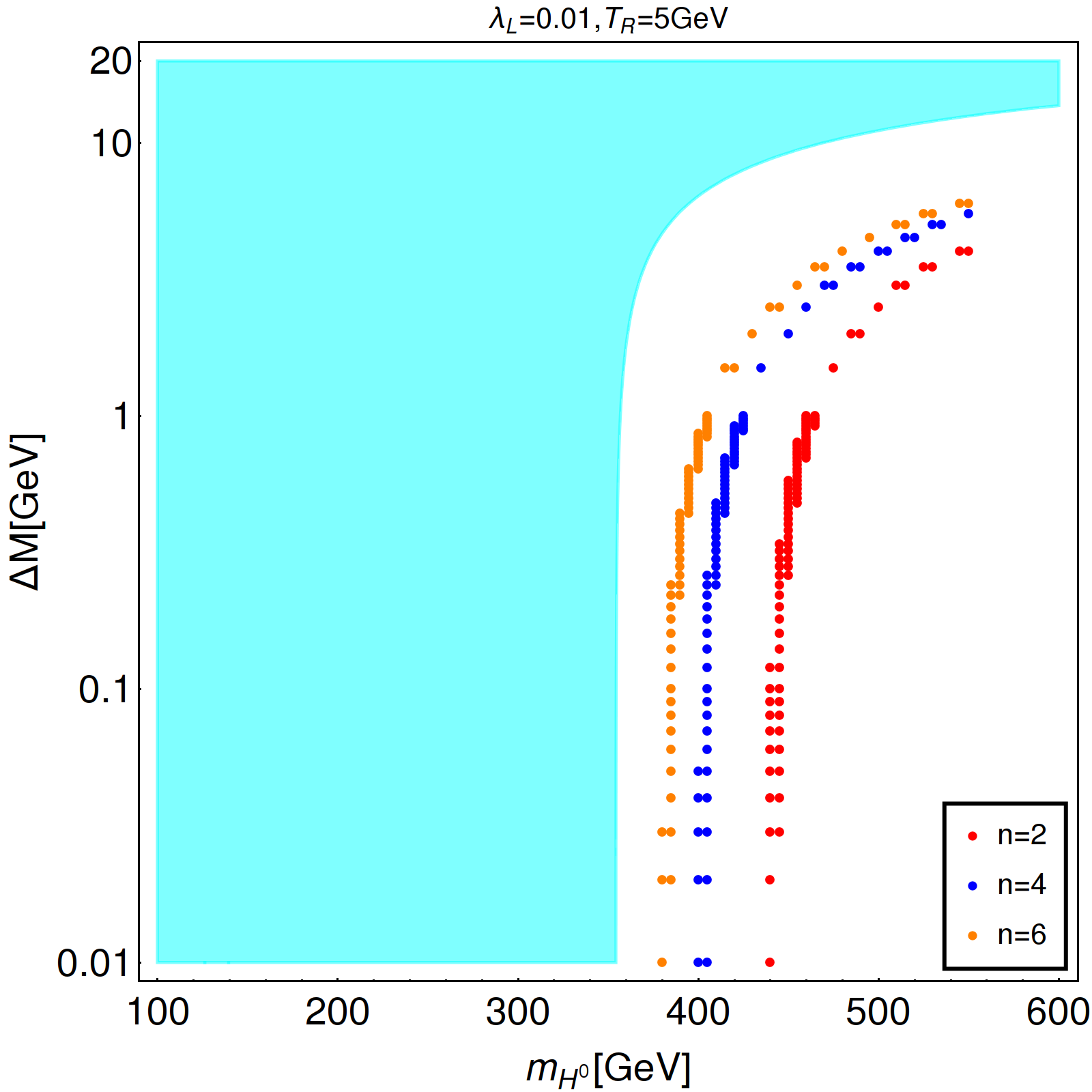}}\quad
  \caption{Relic satisfied points (red, blue, orange) are shown in $m_{H^0}-\Delta M$ plane as function of $n$ values considering (a) $T_R=3$ GeV, (b)$T_R=4$ GeV, (c)$T_R=5$ GeV and an uniform $\lambda_L=0.01$ value. The  cyan region is forbidden from indirect search bound to $WW$ final state.}\label{fig:param-spaceDMmH0}
\end{figure*}

The Boltzmann equation (BEQ) that governs the evolution of comoving number density of the DM, in the standard radiation dominated Universe has the familiar form~\cite{Kolb:1990vq}  

\begin{align}\label{eq:BoltzDM1}
 \frac{dY_{\rm DM}}{dx}=-\frac{\langle\sigma v\rangle s}{\mathcal{H}_R(T) x}\Bigl(Y_{\rm DM}^2-Y_{\rm DM}^{\rm eq^2}\Bigr),
\end{align}


\noindent where, $x=\frac{m_{H^0}}{T}$ and $\langle\sigma v\rangle$ stands for the thermally averaged annihilation cross section. It is always convenient to re-cast the DM number density in terms of the dimensionless quantity $Y_\text{DM}=n_\text{DM}/s$ with $s$ being the entropy per comoving volume. The equilibrium number density of the DM component, in terms of the yield $Y$ is given by



{
\bea
Y_\text{DM}^\text{eq}= \frac{45}{4\pi^4}\Biggl(\frac{g_\text{DM}}{g_{* s}}\Biggr)x^2 K_2\left(x\right)
\eea
}

\noindent where $K_2\left(x\right)$ is the reduced Bessel function of the second kind. For the fast expanding Universe, $\mathcal{H}_R$ in Eq.~\eqref{eq:BoltzDM1} will be replaced by $\mathcal{H}$ of Eq.~\eqref{eq:mod-hubl} leading to




\bea\begin{aligned}
& \frac{dY_{\rm DM}}{dx}=-\frac{A \langle\sigma v\rangle}{x^{2-n/2}\sqrt{x^n+\left(\frac{m_\text{DM}}{T_R}\right)^n}}\,\Bigl(Y_{\rm DM}^2-Y_{\rm DM}^{\rm eq^2}\Bigr)    
    \end{aligned}\label{eq:BoltzDM2}
\eea

\noindent with $A=\frac{2\sqrt{2}\pi}{3\sqrt{5}} \frac{g_{* s}}{\sqrt{g_*}} M_\text{pl}m_\text{DM}$. This is the BEQ of our interest.

As clarified before, in presence of the species $\eta$ with $n>0$, the freeze out of DM occurs at earlier times compared to the case for radiation dominated Universe. In post freeze out time the DM number density still keeps decreasing due to faster red-shift of the energy density of $\eta$ and constant attempt of the DM to go back to thermal equilibrium till the Universe reaches radiation domination, and finally the rate of interaction $\Gamma_{\rm DM}\ll \mathcal{H}_R$. The rate of decrease of the DM relic abundance in this phase is rapid for larger $n$. An approximate analytical solution for the DM yield considering $s$- wave annihilation in this regime reads

\bea\begin{aligned}
& Y_\text{DM}\left(x\right)\simeq\left\{
                \begin{array}{ll}
                 \frac{x_r}{A\langle\sigma v\rangle}\Biggl[\frac{2}{x_f}+\log\left(x/x_f\right)\Biggr]^{-1},~~(n=2)\\
                  \frac{x_r^{n/2}}{2A \langle\sigma v\rangle}\Biggl[x_f^{n/2-2}+\frac{x^{n/2-1}-x_f^{n/2-1}}{n-2}\Biggr]^{-1}~~(n\neq 2)\\
                  \end{array}
              \right.     
    \end{aligned}\label{eq:mod-yld}
\eea

\noindent as reported in Appendix.~\ref{sec:analyt-yld} with $x_{f(r)}=m_\text{DM}/T_{f(R)}$. 
It is evident from Eq.(\ref{eq:mod-yld}), for $n=2$ after freeze-out one can observe the slow logarithmic decrease (although faster than the usual scenario) in the DM yield. The slow logarithmic decrease in the number density is the result of the relentless effort of the DM to go back to the thermal equilibrium\footnote{This feature has been referred to as `relentless' DM in~\cite{DEramo:2017gpl}}. This behaviour continues till $T\simeq T_R$ after which the Universe becomes radiation dominated and the DM comoving number density attains a constant value. For $n>2$ the effect of fast expansion is even more pronounced as the DM yield has a pure power law dependence instead of a logarithmic one. Same as before, the DM number density keeps decreasing until radiation takes over. Similar behaviour can be seen for $p$-wave annihilation as elaborated in~\cite{DEramo:2017gpl}. For different choices of the relevant parameters we shall solve Eq.~\eqref{eq:BoltzDM2} numerically to obtain the DM relic abundance via

\bea\Omega_{\rm DM} h^2= 2.82\times 10^8~m_{H^0} Y_\text{DM}\left(x=\infty\right).\label{eq:dm-relic}
\eea

\noindent This brings us to the independent parameters for IDM dark matter model in a fast expanding Universe that is going to affect the DM relic abundance:

\begin{align}
\left\{m_{H^0},\Delta M, \lambda_L, n, T_R\right\}.
\end{align}

\noindent Note that the presence of last two parameters are due to consideration of fast expansion.


Apart from the requirement of obtaining the PLANCK observed relic abundance ($\Omega_{\rm DM} h^2 = 0.120\pm 0.001 $ at 90$\%$ CL \cite{Aghanim:2018eyx}), there are two other sources that impose severe bound on the IDM {\it desert} region. The spin-independent direct search puts a stringent bound on the IDM parameter space by constraining the DM-nucleon direct detection cross-section. At the tree-level the DM-nucleon scattering cross-section mediated by the SM-like Higgs boson reads~\cite{Cline:2013gha}

\begin{align}
\sigma_{n-H^0}^\text{SI}=\frac{\lambda_L^2 f_N^2}{\pi}\frac{\mu^2 m_n^2}{m_h^4 m_{H^0}^2},\label{eq:DD}
\end{align}

\noindent where $f_N = 0.2837$ represents the form factor of nucleon, $m_n = 0.939$ GeV denotes the nucleon mass and $\mu=m_n m_{H^0}/\left(m_n+m_{H^0}\right)$ is the DM-nucleon reduced mass. The spin-independent direct search exclusion limit puts bound on the model parameters, especially on the coupling $\lambda_L$ and DM mass $m_{H^0}$ via Eq.~\eqref{eq:DD}, which in turn restricts the relic density allowed parameter space to remain viable within the direct search limit. In our work we shall consider the recent XENON1T bound~\cite{Aprile:2018dbl} to restrict the parameter space wherever applicable.


The second most rigorous bound arises from the indirect search experiments that look for astrophysical sources of SM particles produced through DM annihilations or via DM decays. Amongst these final states, the neutral and stable particles {\it e.g.,} photon and neutrinos, can reach indirect detection detectors without getting affected much by intermediate regions. If the emitted photons lie in the gamma ray regime, that can be measured at space-based telescopes like the Fermi-LAT~\cite{Fermi-LAT:2016uux} or ground-based telescopes like MAGIC~\cite{Ahnen:2016qkx}. Now it turns out that for single component IDM candidate, the indirect search severely restricts the thermal average cross section of $H^0 H^0\rightarrow W^+ W^-$ annihilation process. Since bound on other annihilation processes of IDM DM candidate are comparatively milder, we shall mostly focus into the bound on $W^+W^-$ final states for constraining the parameter space. Equipped with these we now move on to investigate the fate of the IDM {\it desert} under the influence of fast expansion.

\subsubsection{IDM dark matter yield in fast expanding background}\label{sec:dm-yld}

As stated earlier, we work in the standard freeze-out regime where we solve Eq.~\eqref{eq:BoltzDM1} with the assumption that the DM was in thermal equilibrium in the early Universe. In order to illustrate the effect of modified BEQ on the DM abundance we deliberately consider a few benchmark values such that they provide under abundance in case of standard Universe (RD), thus falling into the {\it {\it desert}} region. Before delving into the parameter scan we would first like to demonstrate the effect of fast expansion {\it i.e.,} the parameters $\{n,T_R\}$ on the DM yield. In order to do that we fix the coupling $\lambda_L=0.01$ and choose several values of $\{m_{H^0},\Delta M,n,T_R\}$ and obtain resulting DM yield by solving Eq.~\eqref{eq:BoltzDM2} numerically as stated earlier. As we shall see later such a choice of $\lambda_L$ keeps the DM safe from spin-independent (SI) direct search exclusion limits. We have used the publicly available code {\tt micrOmegas}~\cite{Belanger:2010pz} for obtaining the annihilation cross-section $\langle\sigma v\rangle$ and fed them to the modified BEQ in Eq.~\eqref{eq:BoltzDM2} to extract the DM yield.

\begin{figure*}[htb!]
  \centering
  \subfigure[]{\includegraphics[scale=0.32]{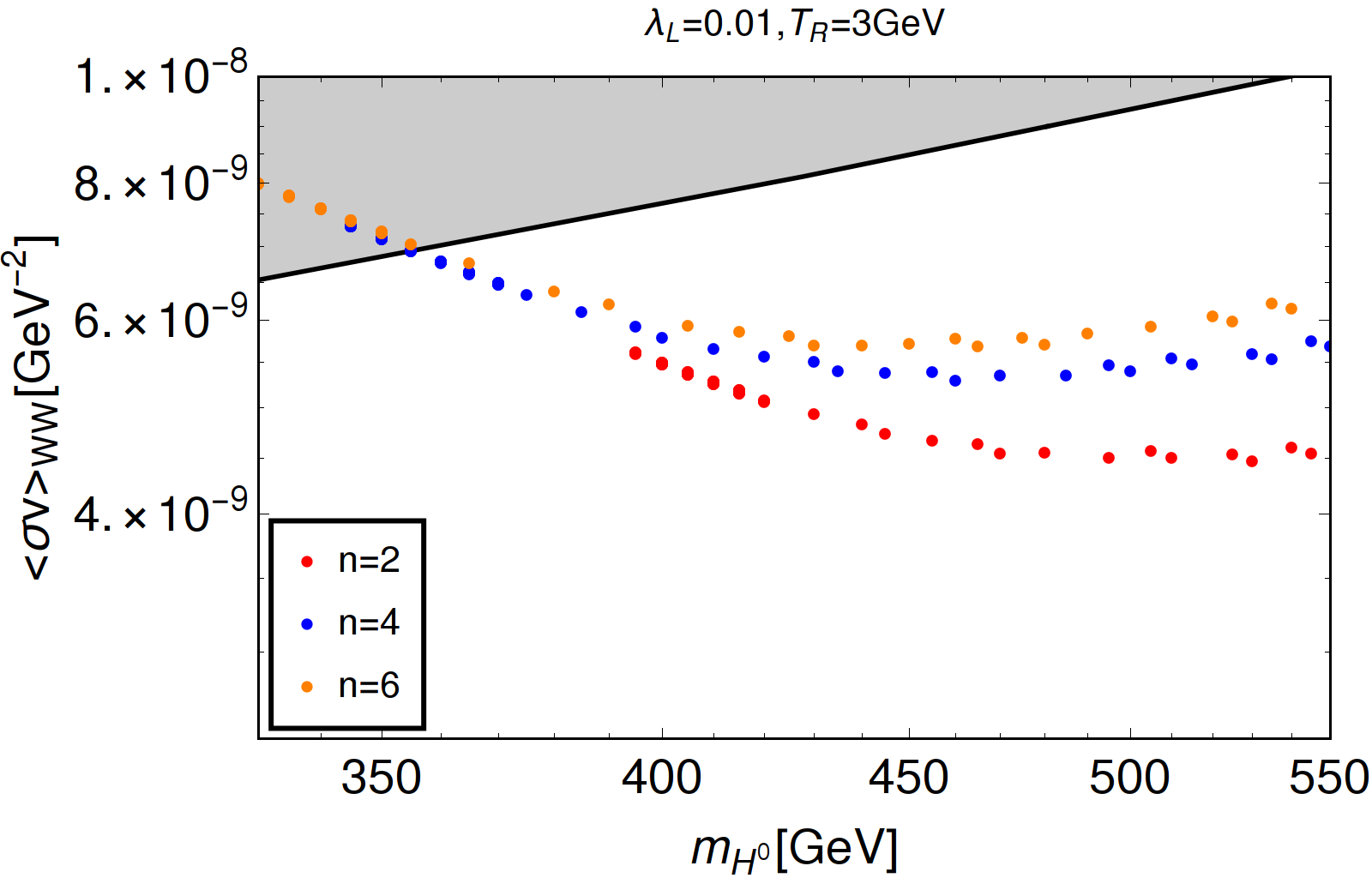}}\quad
  \subfigure[]{\includegraphics[scale=0.32]{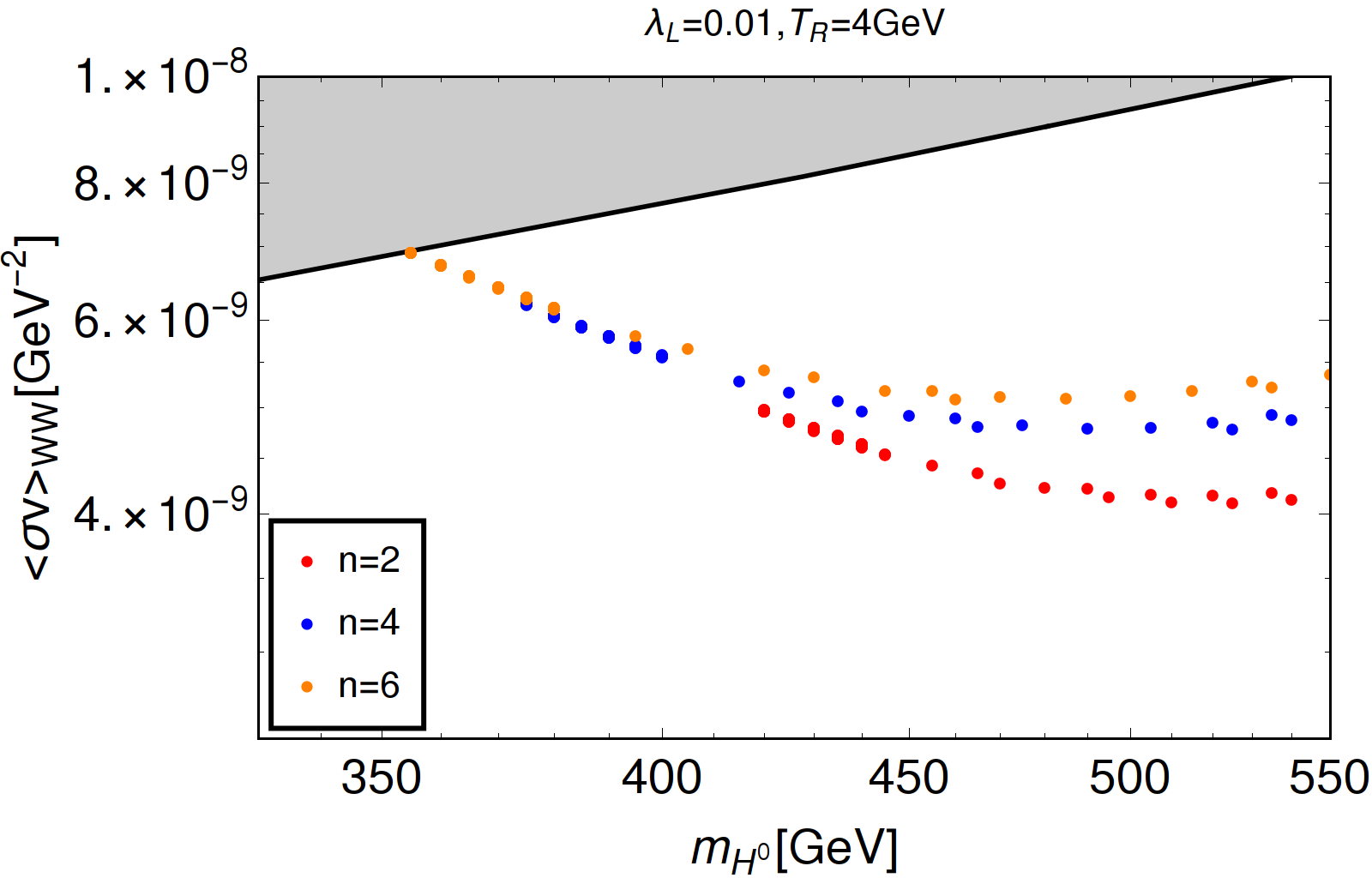}}\quad
  \subfigure[]{\includegraphics[scale=0.32]{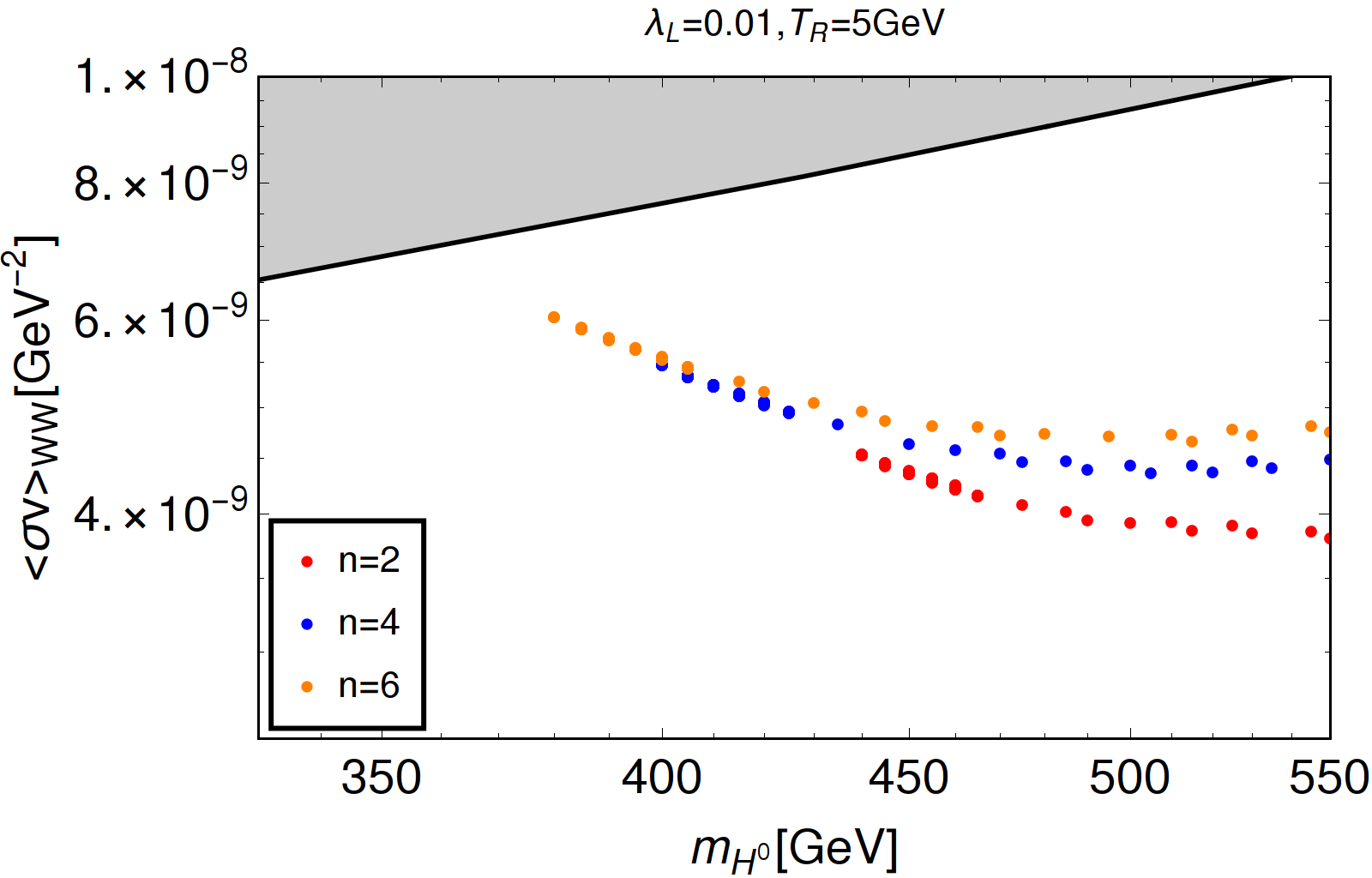}}\quad
  \caption{Numerical estimate of DM annihilation cross section to $W^+W^-$ final states for the relic satisfied points with different $n$ values as shown in Fig. \ref{fig:param-spaceDMmH0} considering (a) $T_R=3$ GeV, (b)$T_R=4$ GeV, (c)$T_R=5$ GeV and an uniform $\lambda_L=0.01$ value. The black solid line represents the latest bound of non-observation of the DM at Fermi experiment.}\label{fig:param-spaceSigmDM}
\end{figure*}

\begin{itemize}
\item For the benchmark values $m_{H^0}=300~\rm GeV$ and $\Delta M=1~\rm GeV$ we fix $T_R=3$ GeV. In the left panel of Fig.~\ref{fig:dm-relic}, we show the evolution of DM abundance as function of $x=m_{H^0}/T$. The solid red colored curve is the case of standard RD Universe $(n=0)$ that clearly shows the DM relic is under abundant for the chosen benchmark. As we increase the value of $n$ from zero, the final relic abundance gets enhanced obeying Eq.~\eqref{eq:dm-relic} and for $n=2$ the relic abundance is satisfied. This typical behaviour surfaces because of the presence of fast expanding component $\eta$ during the DM freeze out. Since the Hubble is larger than that in the RD Universe, the DM freezes out earlier and causes an over production of relic that can be tamed down by suitable choice of the free parameters $\{n,T_R\}$. Needless to mention that $n\to 0$ for a fixed $T_R$ simply reproduces the RD scenario with the unmodified Hubble rate.

\item Next, in the right panel of Fig.~\ref{fig:dm-relic} we fix $n=4$ for the same DM mass of $m_{H^0}=300~\rm GeV$ and choose different values of $T_R$. As one can see from the left panel of Fig.~\ref{fig:dm-relic} $n=4$ corresponds to DM over abundance for $T_R=3~\rm GeV$ (shown by the green dashed curve). In order to obtain the right abundance for $n=4$ one then has to go to a larger $T_R$ obeying Eq.~\eqref{eq:dm-relic} to tame down the Hubble rate. This is exactly what we see here. The correct DM abundance is achieved for $T_R=5~\rm GeV$ with $n=4$. Increase $T_R$ further shall make the DM under abundance as $T_R\to\infty$ for a fixed $n$ corresponds to the standard RD scenario.

\end{itemize}

We thus see the general trend here that when we invoke fast expansion through the Hubble parameter then for certain choices of $\{n,T_R\}$ it is indeed possible to revive the region of the DM parameter space that is otherwise under abundant (shown by the red curve in each plot). Our next task is to see the relic density allowed parameter space that survives once direct and indirect search bounds are imposed. 

Before we proceed, it is necessary to check whether the DM ever thermalizes in the fast expanding  Universe at some early time {validating the BEQ (Eq.~\eqref{eq:BoltzDM2}) that we are using to find its yield}. Thermalization can be accomplished by satisfying the condition $\Gamma_\text{int} > \mathcal{H}(T)$ at some high temperature above the weak scale ($\sim \mathcal{O}(1) {\rm ~TeV}$). Considering the temperatures larger than the DM mass, the scattering rate of the DM can be approximated as~\cite{DEramo:2017gpl} 

\begin{align}
 \Gamma_\text{int}=n_\text{DM}\langle\sigma v\rangle\simeq\frac{\zeta(3)T^3}{2\pi^2}\frac{g_2^4}{32\pi}\frac{T^2}{(T^2+M_{\rm med}^2)^2} \label{eq:non-therm},
\end{align}

\noindent where $g_2$ is the $SU(2)_L$ gauge coupling and $M_{\rm med}$ is the mediator mass \footnote{For point interaction we can consider $M_\text{med}\to 0$.}. For the inert doublet model, in principle $\lambda_L$ (one of the scalar couplings) should also enter into the Eq.(\ref{eq:non-therm})  Now, since $\lambda_L\ll g_2$ (motivated from satisfying the direct search bound) always holds in our analysis, we have found that the DM pair annihilation is always dominated by the gauge boson final state which is proportional to the coupling strength $g_2^4$~\cite{LopezHonorez:2006gr,LopezHonorez:2010tb}. In Fig. \ref{fig:Oeq-Approx}, we compare the modified Hubble rate with the DM interaction rate as function of temperature $T$, considering $T_R=1$ GeV for different values of $n$. In the left panel of Fig. \ref{fig:Oeq-Approx} we consider the approximate analytical relation in Eq.~\eqref{eq:non-therm}, while for the right panel we calculate the DM interaction rate numerically by evaluating the annihilation cross-section using {\tt micrOMEGAS}~\cite{Belanger:2008sj} for a DM of mass $m_{H^0}=350$ GeV. We notice, the approximate expression closely follows the numerically obtained result, {implying the annihilation rate of DM is largely independent of its mass}. From these plots we see, for $n=6$, thermalization is achieved at temperature $T\gtrsim 2.5~\rm TeV$ for $T_R=1$ GeV. For $T_R>1~\rm GeV$ the DM thermalizes much earlier (at a larger temperature) as the modified Hubble rate decreases following Eq.~\eqref{eq:mod-hubl} and it could allow higher $n>6$ values. The same conclusion can be drawn for the case of inert triplet DM  where the dominant annihilation channel is again due to gauge boson final states, and hence determined by the $SU(2)_L$ gauge coupling. In case where the DM interaction rate is always below the Hubble rate, the thermal production of the DM is not possible, and we need to opt for the non thermal case with modified Boltzmann equations. Taking thermalization of the DM in the early Universe into account, we confine ourselves within the range $2\leq n\leq 6$ with $T_R\gtrsim 1~\rm GeV$, unless otherwise mentioned explicitly \footnote{Lowering $T_R$ $(< 1~\rm GeV)$ disallows higher $n$-values from the requirement of thermalization above the weak scale.}. We find, within the said range of $n,T_R$, both inert doublet and inert triplet DM achieve thermal equilibrium for the mass range $m_\text{DM}\lesssim 525~\rm GeV$ and $m_\text{DM}\lesssim 1.9~\rm TeV$ respectively, at a temperature above the weak scale.


\subsubsection{Allowed parameter space for IDM dark matter in a fast expanding Universe}\label{sec:dm-param-scan}
\begin{figure*}[htb!]
  \centering
  \subfigure[]{\includegraphics[scale=0.28]{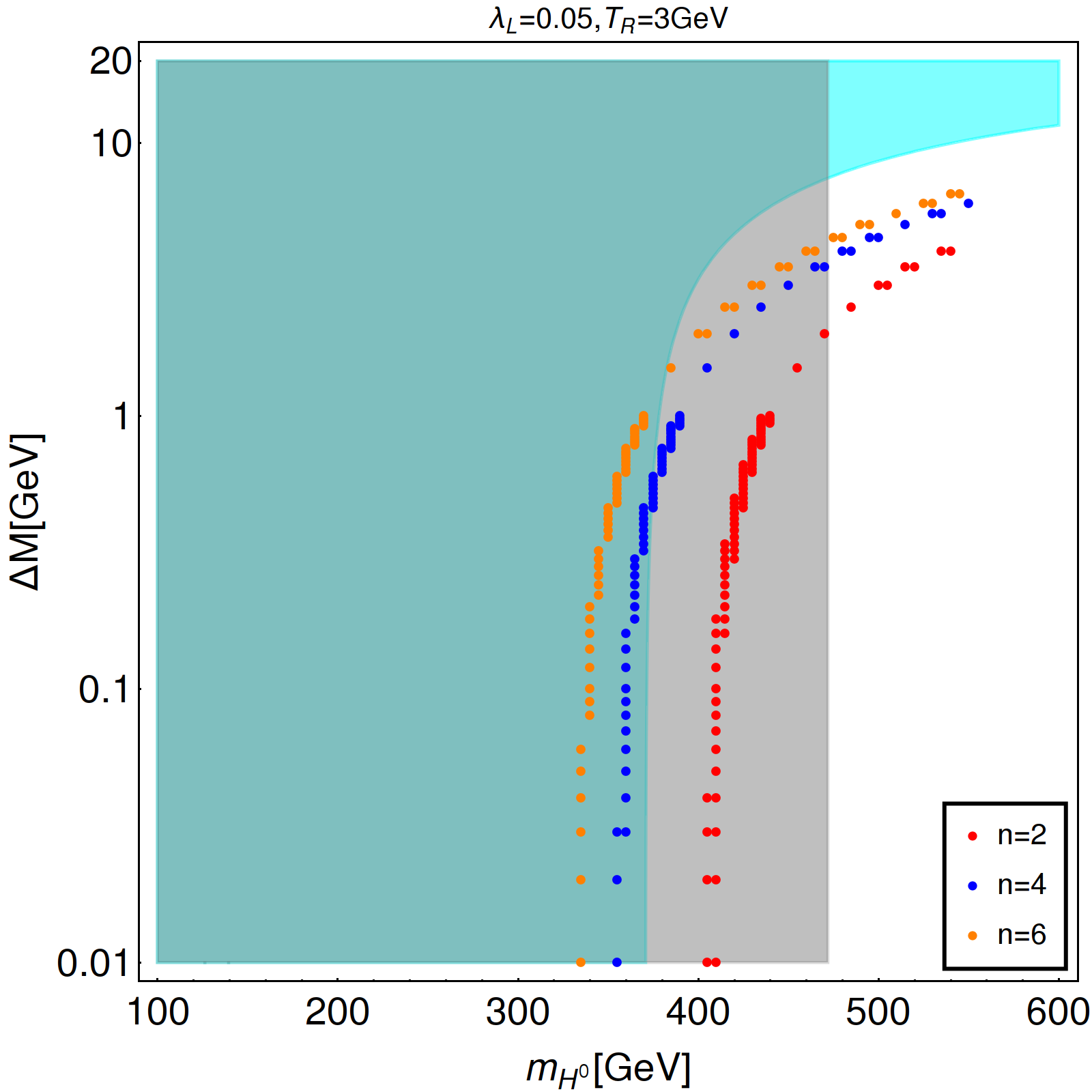}}\quad
  \subfigure[]{\includegraphics[scale=0.28]{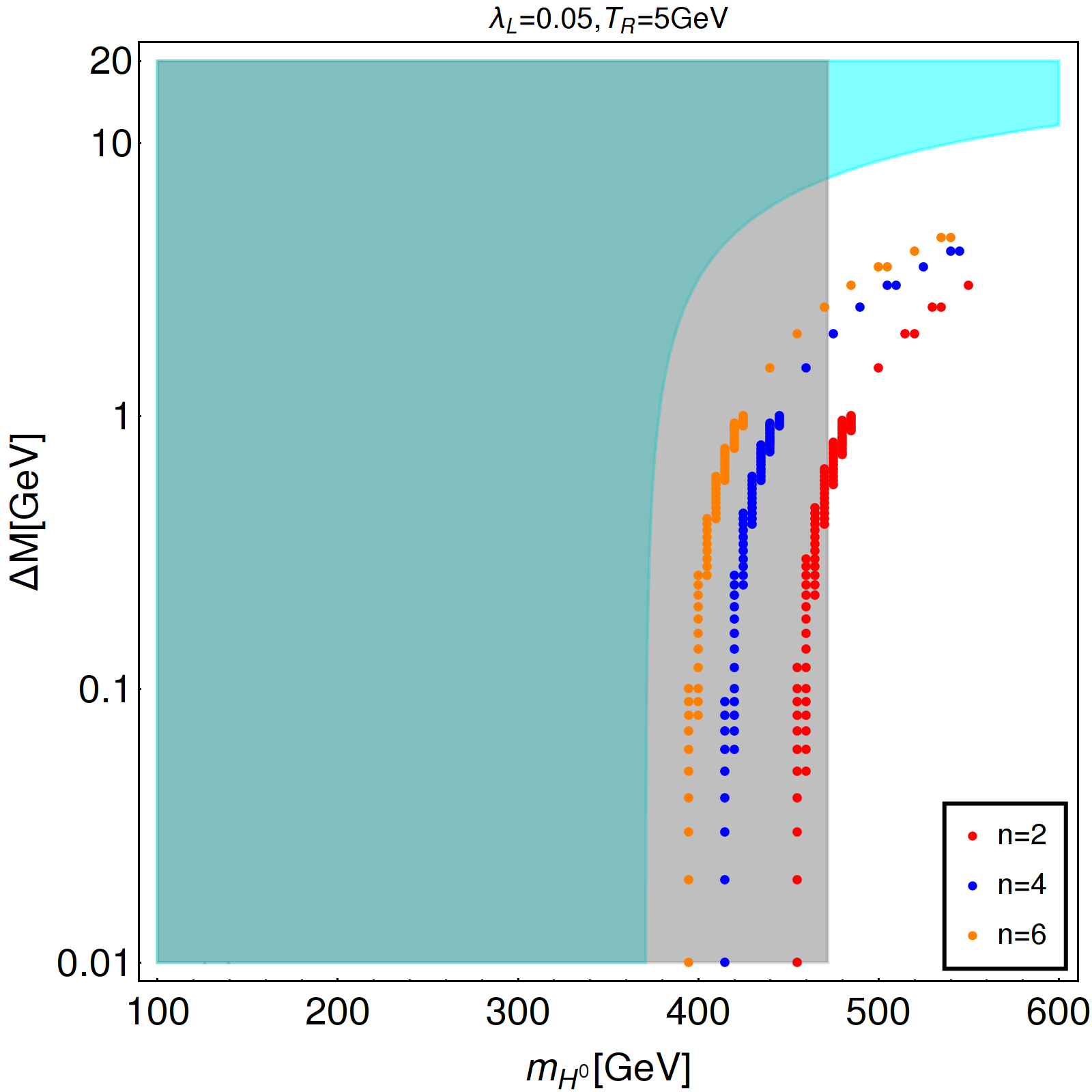}}\quad
  \subfigure[]{\includegraphics[scale=0.28]{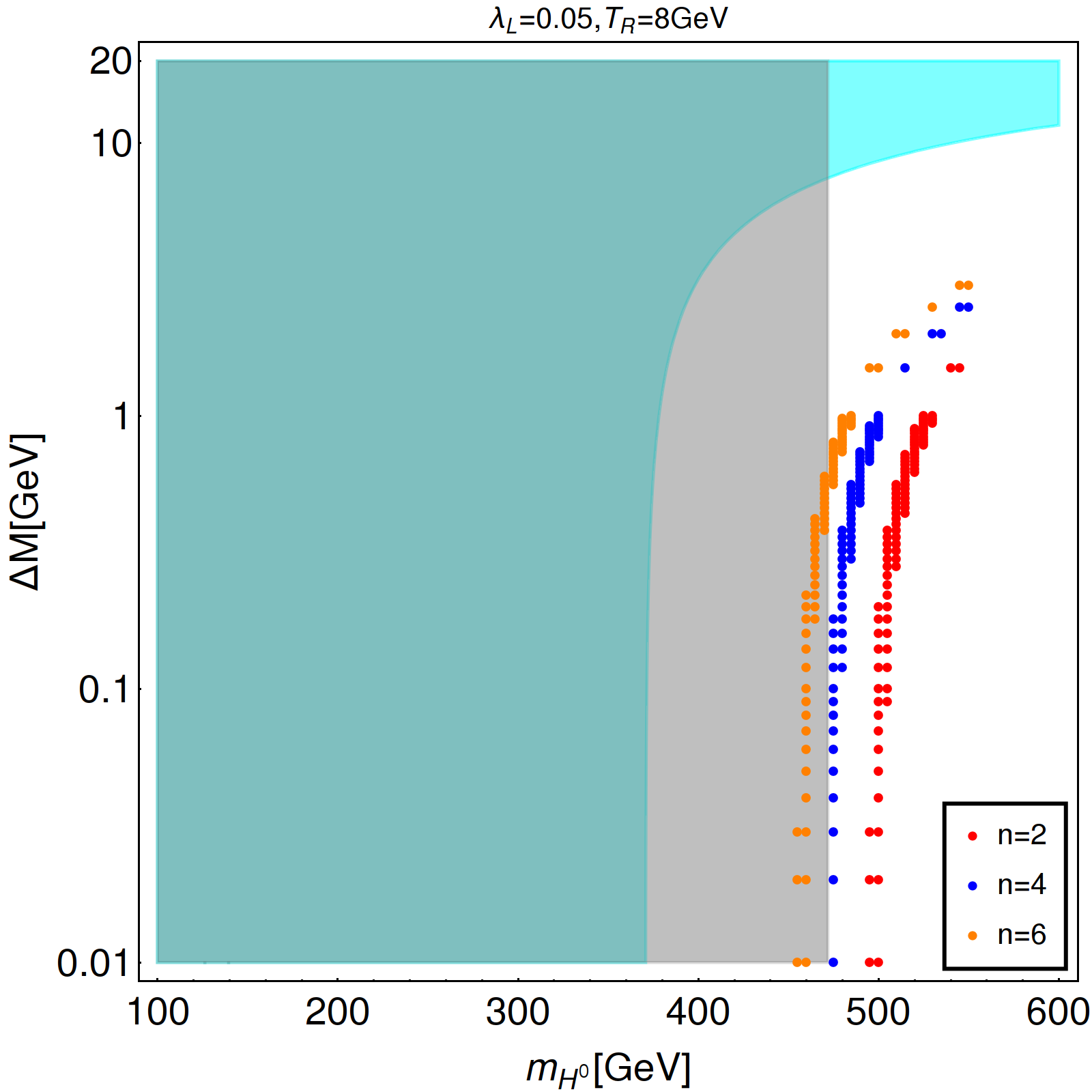}}\quad
  \caption{Relic satisfied points (red, blue, orange) are shown in $m_{H^0}-\Delta M$ plane as function of $n$ values considering (a) $T_R=3$ GeV, (b)$T_R=5$ GeV, (c)$T_R=8$ GeV and an uniform $\lambda_L=0.05$ value. The indirect search bound for $W^+W^-$ final state forbids the cyan region while the gray shaded region shows direct search exclusion limit from XENON1T.}\label{fig:param-spaceDMmH02}
\end{figure*}

To find out how much of the relic density allowed parameter space is left in a fast expanding framework after satisfying (in)direct detection bounds we would like to perform a scan over the relevant parameter space. In order to do that, first we fix $\lambda_L=0.01$ as before. In that case the remaining parameters relevant for DM phenomenology are $\{m_{H^0},\Delta M, n,T_R\}$. In Fig.~\ref{fig:param-spaceDMmH0} we have shown the relic satisfied points in $\Delta M-m_{H^0}$ plane by varying $n$ considering $T_R=\{3,4,5\}$ GeV. The cyan shaded region violates the indirect detection bounds from Fermi-LAT $WW$ final state, and hence forbidden. For a constant $\Delta M$ and $T_R$ notice that a larger value of $n$ requires smaller DM mass to satisfy the relic bound. This particular nature appears since larger $n$ leads to enhanced expansion rate of the Universe and hence the DM annihilation rate should be sufficient enough to avoid early freeze out and subsequently over-abundance. Thus, a smaller value of $m_{H^0}$ is preferred to be within the relic limit since the annihilation rate of the DM goes roughly as $\langle\sigma v\rangle\propto 1/m_{H^0}^2$. However such requirement of enhanced annihilation cross section due to larger $n$ may get disfavored by the indirect search bound as one can see in the left most panel of Fig.~\ref{fig:param-spaceDMmH0}. This can be evaded if we increase the $T_R$ as well, since then it reduces the Hubble rate against larger $n$ following Eq.~\ref{eq:mod-hubl}. Such pattern can be observed in the other two figures for $T_R=\{4,5\}~\rm GeV$. The bound arising from spin-independent direct detection cross section for $\lambda_L=0.01$ is weak and does not appear in Fig.~\ref{fig:param-spaceDMmH0}. 

For a clear insight on the detection prospect of the DM at indirect detection experiments, in Fig.~\ref{fig:param-spaceSigmDM}, we estimate the numerical values of $\langle \sigma v\rangle$ for $W^+ W^-$ final states of the relic satisfied points as shown earlier in Fig.~\ref{fig:param-spaceDMmH0}. The latest exclusion bound from Fermi experiment due to non-observation of DM signal has been shown via the solid black line. In accordance with the earlier trend it can be seen that increasing $T_R$ reduces the $\langle \sigma v\rangle$ for a particular $n$. Hence improved sensitivity of the Fermi experiment has the ability to probe or rule out the cases particularly with low $T_R$ values. So far we have worked with $\lambda_L=0.01$. We would now like to see the consequence of a relatively larger $\lambda_L=0.05$ on the DM phenomenology. As it is evident from Eq~\eqref{eq:DD}, the direct detection cross-section becomes important for a larger $\lambda_L$. In Fig.~\ref{fig:param-spaceDMmH02} we present the relic satisfied points in the bi-dimensional plane of $M_{H^0}-\Delta M$ for different sets of $\{n,T_R\}$ values. As expected, we find that for $\lambda_L=0.05$ the spin independent direct detection constraints become dominant over the indirect detection ones in the mass region $m_{H^0}\lesssim480~\rm GeV$. The characteristics of relic satisfied contours are same as those portrayed for the case with $\lambda_L=0.01$ corresponds lower value of $T_R$ with other parameters are fixed. As we see, for larger $T_R$ and smaller $n$, the relic satisfied points with $0.01\lesssim\Delta M\lesssim 10~\rm GeV$ are unconstrained from both direct and indirect detection. More precisely, $\Delta M<3~\rm GeV$ is ruled out for $T_R=3~\rm GeV$ and $n=4$, but on increasing $T_R$ to $8~\rm GeV$, the bound on $\Delta M$ is significantly relaxed for DM mass in the same range with the same choices of $n$. 

\begin{figure*}[htb!]
  \centering
  \subfigure[]{\includegraphics[scale=0.32]{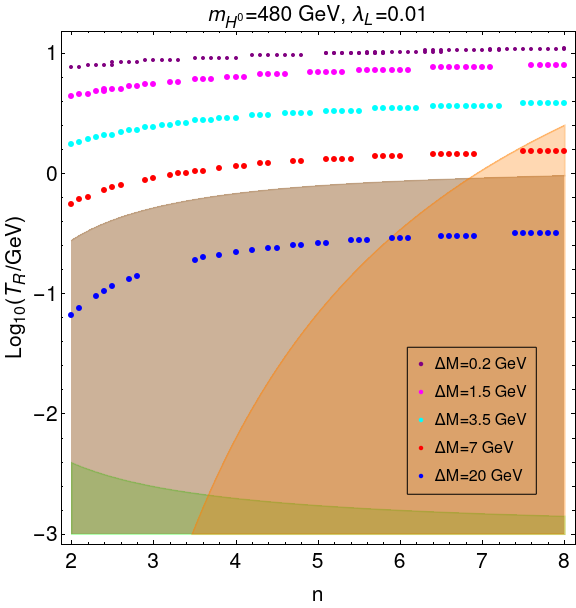}}\quad~~~~
  \subfigure[]{\includegraphics[scale=0.32]{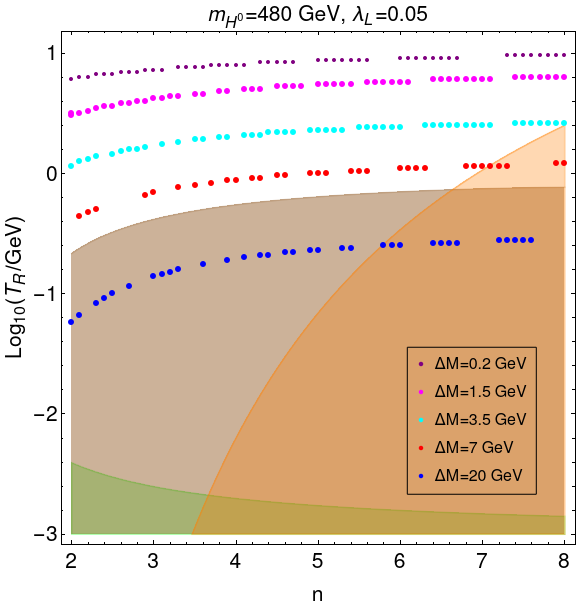}}\quad
  \caption{Relic satisfied points are shown in $n-T_R$ plane for a fixed DM mass of 480 GeV and different choices of $\Delta M$ for (a) $\lambda_L=0.01$ and (b) $\lambda_L=0.05$. In both the plots the green region is forbidden from the BBN bound on $T_R$ following Eq.~\eqref{eq:tr-bbn} while the orange and the brown regions are disallowed by the non-thermalization of DM above weak scale (following Eq.~\eqref{eq:non-therm}) and indirect search constraints respectively. Any point in the $n-T_R$ plane is also subject to additional constraint arising from the perturbative unitarity bound (discussed later) which is relatively weaker.}\label{fig:nTRrelic}
\end{figure*}

\begin{figure*}[htb!]
  \centering
  \subfigure[]{\includegraphics[height=6cm,width=8cm]{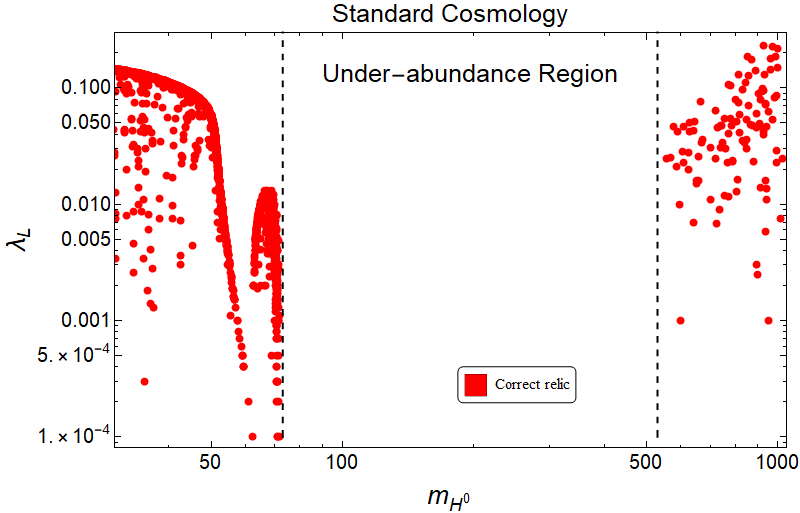}}\quad~~~~
  \subfigure[]{\includegraphics[height=6cm,width=8cm]{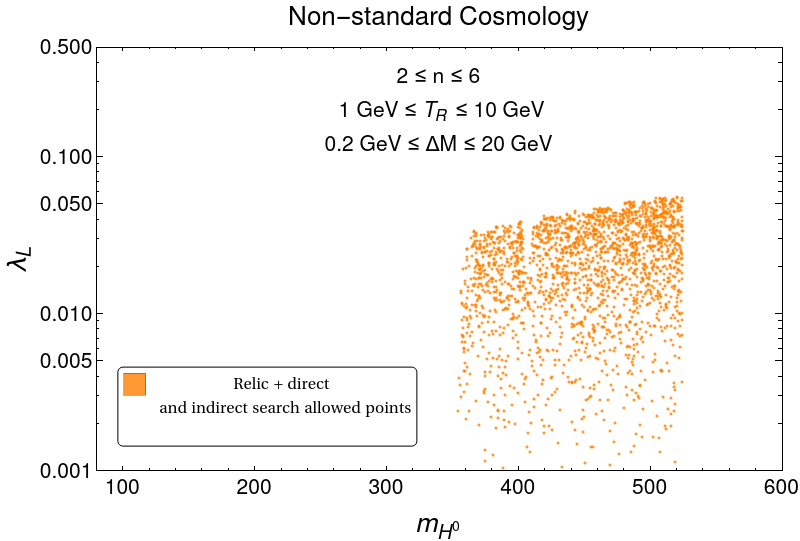}}\quad
  \caption{Left: The IDM parameter space in $m_{H^0}-\lambda_L$ plane validated with PLANCK observed relic density bound considering standard cosmology. The presence of the {\it desert} (relic under-abundant) region for $80 {\rm~ GeV}\lesssim m_{H^0}\lesssim 525$ GeV can clearly be seen. Right: Same as left but in a non-standard cosmological background where the {\it desert} region has been revived satisfying all constraints: relic density, direct detection due to XENON1T and indirect search. The values of the relevant parameters are mentioned in the plot legends.} 
  \label{fig:nTRrelicI}
\end{figure*}

So far we have worked with some discrete values of $\{n, T_R\}$ with $T_R\gtrsim 1$ GeV and $2\leq n \leq 6$. The explicit dependence of the DM relic on the fast expansion parameters $n,T_R$ are shown in Fig.~\ref{fig:nTRrelic} for a fixed DM mass of 480 GeV considering two different values of $\lambda_L\sim $ 0.01 and 0.05. Following the previous scans here we see similar trend for $\lambda_L=\{0.01,0.05\}$ in the left and right panel respectively. With the increase in $\Delta M$ we see a smaller $T_R$ is required to satisfy the observed relic abundance. This can be understood from the fact that a larger $\Delta M$ leads to under abundance (since DM annihilation dominates over the co-annihilation in the given range of $\Delta M$) and hence a smaller $T_R$ is required to trigger a faster expansion following Eq.~\eqref{eq:mod-hubl} to satisfy the DM abundance. For the same reason larger $\Delta M$ requires larger $n$'s for a fixed $T_R$ to produce right relic. Recall that smaller value of $T_R$ for a fixed $n$ (and vice-versa) violating the limit in Eq.~\eqref{eq:tr-bbn} are disfavoured from the BBN bound. This BBN-excluded region (green) is shown in either of the plots in green. Larger $\Delta M (\gtrsim 20)$ GeV regions, as they require smaller $T_R$ to satisfy the relic abundance, get discarded from the BBN bound. The brown region indicates the disallowed space by indirect search constraint which is also present in Figs. \ref{fig:param-spaceDMmH0} and \ref{fig:param-spaceDMmH02} (shown in cyan) while the orange region is disfavored by the violation of DM thermalisation condition before weak scale following Eq.(\ref{eq:mod-hubl}) and Eq.(\ref{eq:non-therm}). 

In principle a lower bound on $\Delta M$ should also be present in Fig.~\ref{fig:nTRrelic} arising from the condition the heavier mass eigenstates should decay completely before the BBN. However we find that the obtained bound already lies below our working range of $\Delta M$ as specified earlier and hence does not appear in Fig.~\ref{fig:nTRrelic}. We also see that, for fixed $\Delta M$ and $m_{H^0}$, larger $\lambda_L$ prefers low $T_R$ (for a fixed $n$) or larger $n$ (for a fixed $T_R$). This is typically attributed to the DM annihilation cross-section that has a quadratic dependence on $\lambda_L$.  {The requirement of thermalization of the DM above the weak scale disallows larger values of $n$ for smaller $T_R$ as shown by the orange region. A smaller $T_R$ results in a faster expansion causing the DM to fall out of thermal equilibrium in early times. This can be prohibited by tuning $n$ to smaller values such that the DM thermalizes at temperatures above the weak scale. Thus, larger $n$ values are discarded for smaller $T_R$. This bound remains the same for $\lambda_L=0.05$ (shown in the right panel), since the DM annihilation is dominantly controlled by the gauge coupling $g_2$ as discussed earlier in details.} With these outcomes, it is understandable that the fast expansion parameters are well restricted by all the combined constraints irrespective of the value of $\lambda_L$. Finally, it is crucial to note that the indirect search constraint disfavours DM mass $\lesssim 350~\rm GeV$ immaterial of the choice of $\lambda_L$,  eliminating the possibility of resurrecting the low DM mass region satisfying all relevant constraints{\footnote{This implies the {\it desert} region for IDM, taking into account the indirect search bound, typically lies in the range $350\lesssim m_{H^0}\lesssim 525~\rm GeV$ for small $\lambda_L$.}}. This, together with the direct detection bound (important for larger $\lambda_L$), typically rules out the allowed parameter space for a DM mass of 200 GeV that was overlooked in earlier work~\cite{Mahanta:2019sfo}. This can further be verified from Fig.~\ref{fig:nTRrelicI} where in the left panel we present the allowed points from relic density considering standard cosmology in $m_{H^0}-\lambda_L$ plane. It clearly shows presence of a void (under abundant) in the range $80 { \rm ~GeV}\lesssim m_{H^0}\lesssim 525$ GeV. In the right panel, considering a fast expanding Universe, we perform a random scan for different ranges of the relevant parameters and sort out the points satisfying observed relic abundance, indirect search and direct search due to XENON1T exclusion. 
We find, viable parameter space in the said mass range under non-standard scenario {satisfying all relevant bounds}. Also, non-existence of any allowed points for $m_{H^0}\lesssim 350$ GeV confirms our earlier observations\footnote{This lower bound takes into account the thermalization condition.}. 
From the right panel one can notice, for a given DM mass, it is possible to choose $\lambda_L$ as small as 0.001. For such small $\lambda_L(\lesssim 0.01$), the direct search cross section (Eq.~\eqref{eq:DD}) becomes safe from XENON1T exclusion limit, and indirect search provides the most stringent bound on DM mass (see Fig. \ref{fig:param-spaceDMmH0}). In contrast, for a larger $\lambda_L\gtrsim 0.05$, direct search constraint becomes important (see Fig.\ref{fig:param-spaceDMmH02}). The DM annihilation cross-section (or equivalently, the relic abundance), however, is controlled dominantly by the $SU(2)_L$ gauge coupling, while $\lambda_L$ plays a sub-dominant role. Therefore, in the present set-up, it is possible to work with further lower $\lambda_L(\lesssim 0.001$) satisfying all pertinent bounds, without altering the allowed range of DM mass.


\subsubsection{Collider probe of the IDM {\it {\it desert}} region}\label{sec:collider} 

\begin{figure*}[htb!]
\centering
\subfigure[]{\includegraphics[height=6.2cm,width=8cm]{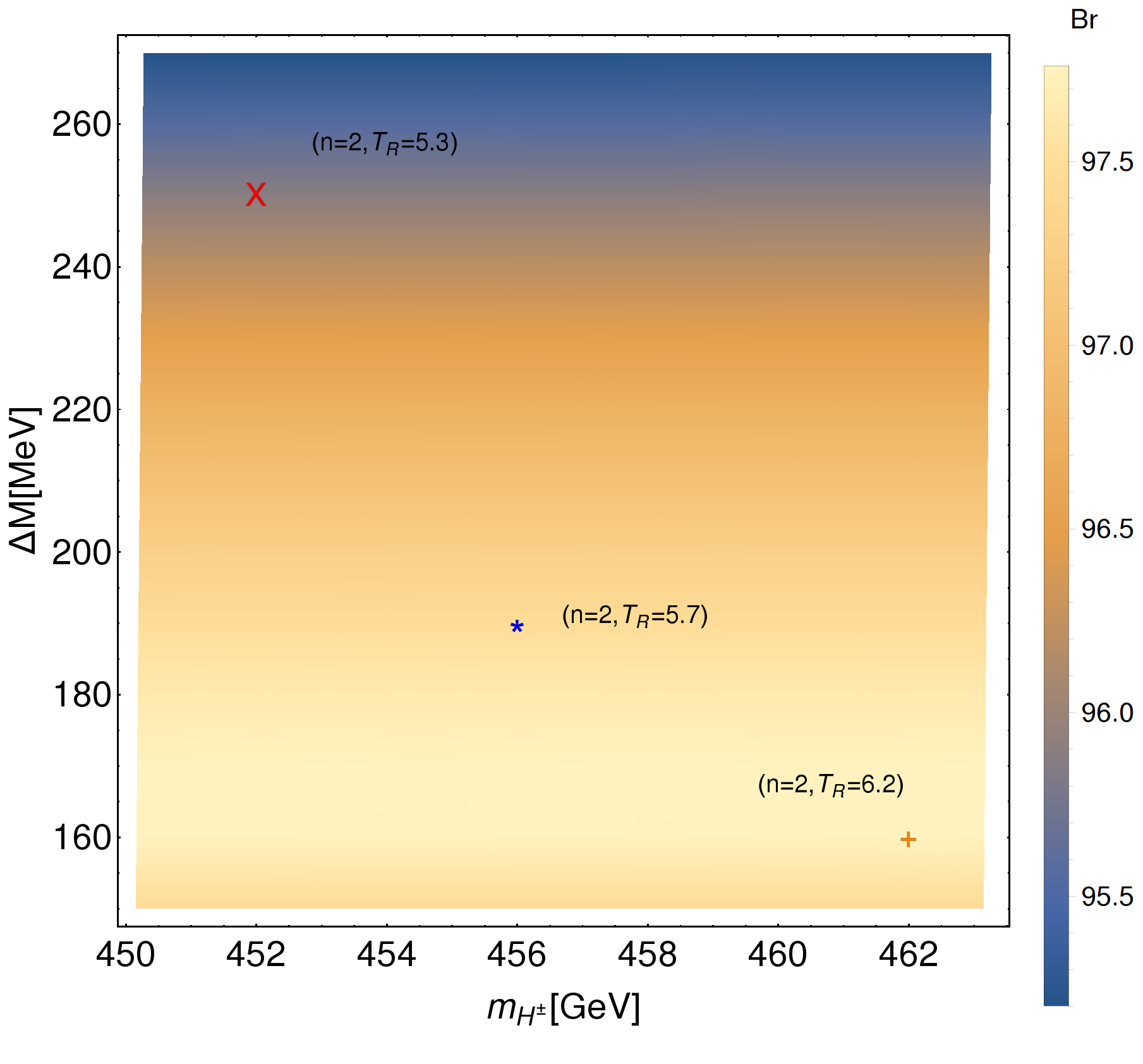}}\quad
\subfigure[]{\includegraphics[height=6.2cm,width=8cm]{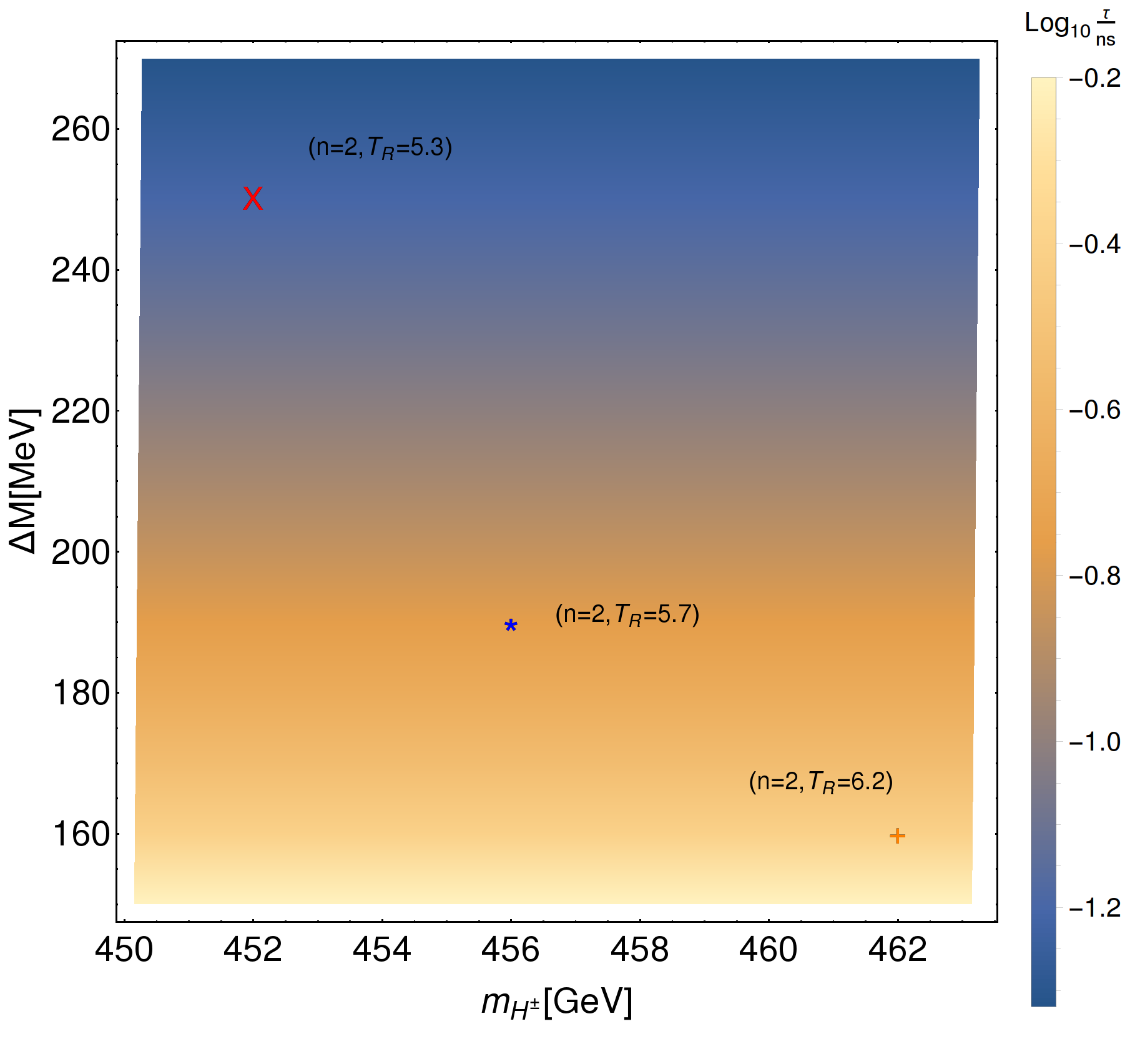}}\quad
\subfigure[]{\includegraphics[height=6.2cm,width=8cm]{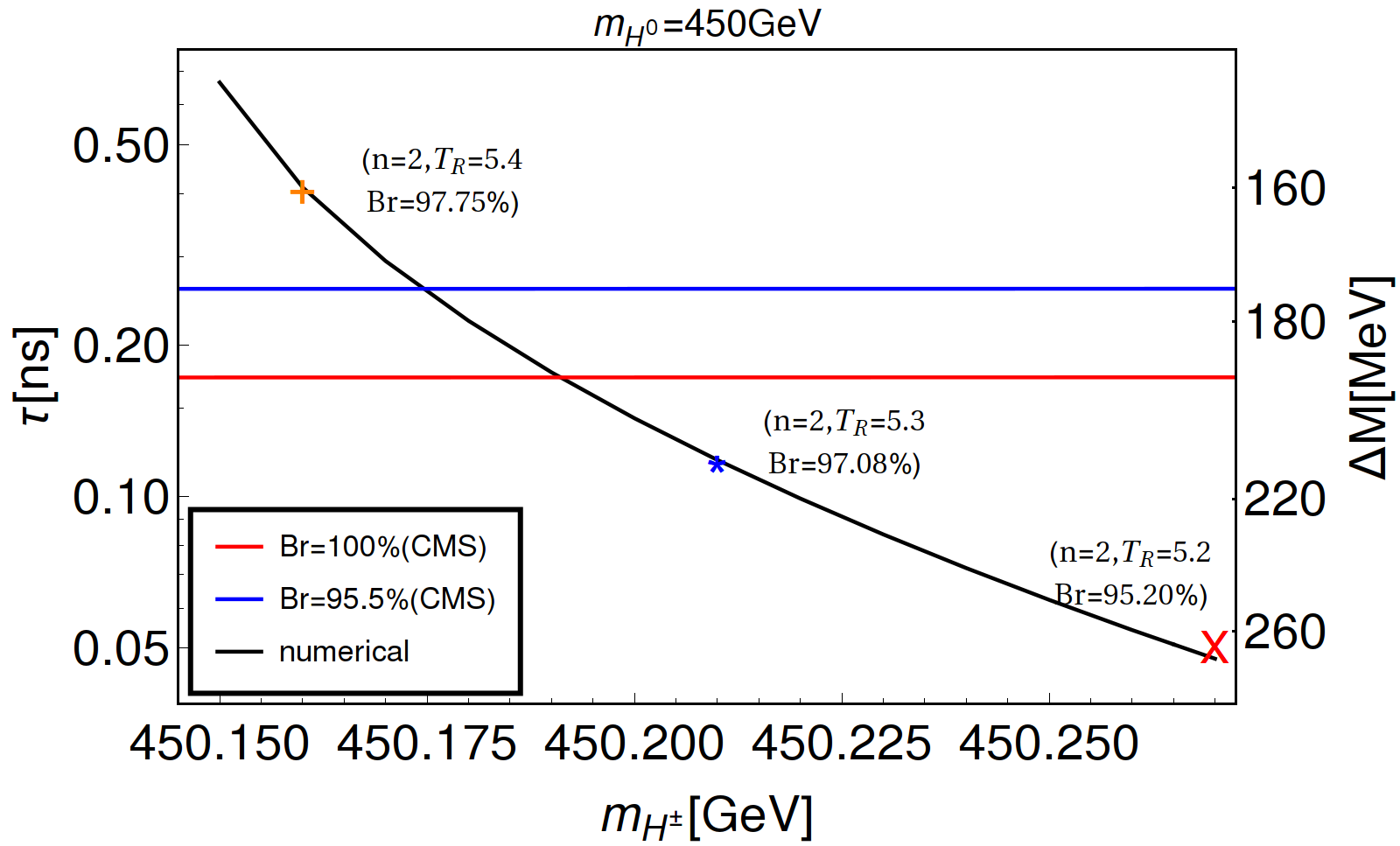}}\quad
\subfigure[]{\includegraphics[height=6.2cm,width=8cm]{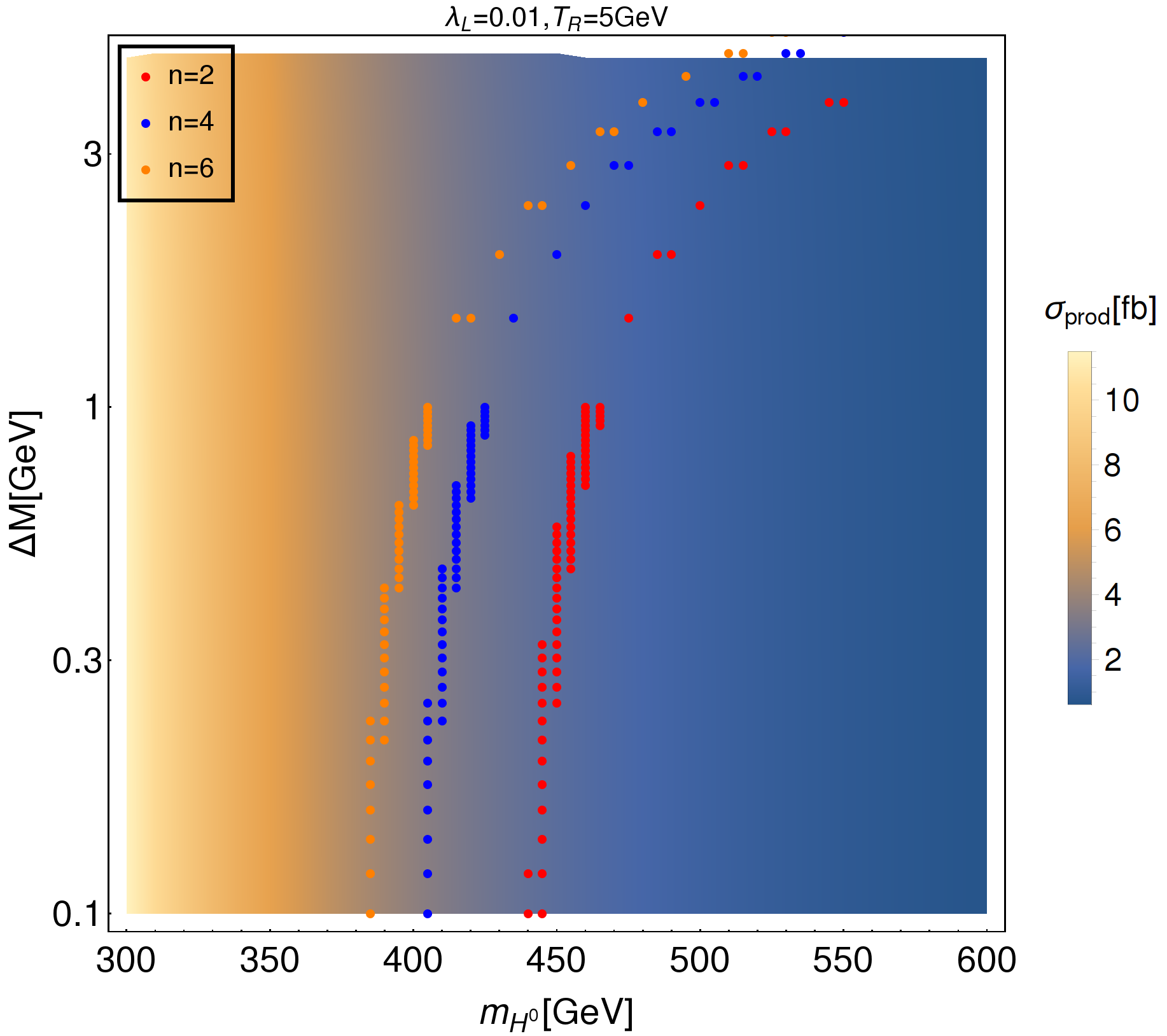}}\quad
\caption{Top Left: Variation of decay branching ratio for $H^\pm$ in the bi-dimensional plane of $m_{H^\pm}-\Delta M$ where the relic density satisfied benchmark points are denoted by ``X", ``$\star$" and ``+" for $n=2$ and different choices of $T_R$ in GeV. Top Right: Same as top left but the variation is shown against the lifetime $\tau$ (in ns) of $H^\pm$ decay. Bottom Left: Total decay lifetime of $H^\pm$ as a function of $m_{H^\pm}$ where the relic satisfied points are marked by ``X", ``$\star$" and ``+" for $n=2$ and different choices of $T_R$ (in GeV). On the same plane we also show exclusion limits from CMS at $\sqrt{s}=13~\rm TeV$ corresponding to 100\% (in red) and 95.5\% (in blue) branching fraction (see text for details). Bottom Right: Variation of production cross-section for $pp\to H^{+}H^{-},H^{\pm}H^0(A^0)$ at $\sqrt{s}=13~\rm TeV$ in the $\Delta M-m_{H^0}$ plane where the colour bar represents the production cross-section in units of $fb$. On the same plane we show DM parameter space allowed by relic density and (in)direct detection for $T_R=5~\rm GeV$ with $n=2,4,6$ in red, blue and orange respectively.}\label{fig:prodcs}
\end{figure*}

As we have already seen, for the mass region of our interest, satisfying relic abundance and exclusion limits from (in)direct searches, the mass splitting $\Delta M$ can be at best a few GeV for any $n\geq0$. Such small $\Delta M$ regions are indeed challenging to probe at the colliders. This extremely compressed scenario can be probed with identifying the charged track signal of a long-lived charged scalar which is $H^\pm$ in this case~\cite{Belyaev:2016lok,Bhardwaj:2019mts}. For $\Delta M\approx 200~\text{MeV}$ the charged scalar has the dominant decay mode: $H^\pm\to\pi^\pm H^0$. Following~\cite{Belyaev:2016lok} one can analytically obtain the $H^\pm\to\pi^\pm H^0$ decay width in the $\Delta M/m_{H^\pm}\ll 1$ limit as:

\bea
\Gamma_{H^\pm\to\pi^\pm H^0} = \frac{g_2^4 f_\pi^2}{64\pi}\frac{\Delta M^3}{m_W^4}\sqrt{1-\frac{m_{\pi^\pm}^2}{\Delta M^2}}\label{eq:pi-life},
\eea

\noindent where $g_2$ is the $SU(2)_L$ gauge coupling strength, the charged pion mass, $m_{\pi^\pm}=139.57$ MeV and $f_\pi\approx 130~\text{MeV}$ is the pion decay constant. Note that the decay width and hence the lifetime $\tau_{H^\pm\to\pi^\pm H^0}\equiv\frac{1}{\Gamma_{H^\pm\to\pi^\pm H^0}}$ of the charged scalar is inversely proportional to the mass splitting. Therefore,  a large mass splitting shall produce a charged track of smaller length and vice versa. {Depending on $\Delta M$ two scenarios can arise:

\begin{itemize}
\item  For $\Delta M\in\{140-200\}~\rm MeV$, $H^\pm$ shall give rise to {\it disappearing} charged track of length $L=c\tau\simeq\mathcal{O}\left(100-10\right)~\rm cm$ { with branching ratio (of $H^\pm\to\pi^\pm H^0$) close to $100\%$. For $\Delta M>200$ MeV the branching ratio gets reduced as new decay modes start to appear.}  

\item For $\Delta M< m_\pi$ the decay is defined via the 3-body process: $H^\pm\to W^\star\,H^0\to \ell\,\nu_{\ell}\,H^0$ which is proportional to $\Delta M^5/m_W^4$. The decay width of such a process turns out to be $\lesssim 10^{-18}~\rm GeV$ resulting in a decay length of $c\tau\gtrsim\mathcal{O}(\rm m)$, implying $H^\pm$ remains stable at collider scales and decay outside the detector giving rise to a {\it stable} charged track.

\end{itemize}

\noindent {We have  used {\tt CalcHEP}~\cite{Belyaev:2012qa} to compute the decay width (total and partial) numerically taking care of both the 2-body and 3-body decay of $H^\pm$.}

A disappearing track results from the decay products of a charged particle which go undetected because they either have too small momentum to be reconstructed or have interaction strength such that they do not produce hits in the tracker and do not deposit significant energy in the calorimeters. Searches for disappearing track signatures have been performed both by CMS~\cite{Sirunyan:2018ldc,CMS-PAS-EXO-19-010} and ATLAS~\cite{Aaboud:2017mpt} in the context of supersymmetry for a center of mass energy of $\sqrt{s}=13~\rm TeV$, setting upper limits on the chragino mass and production cross-section. To recast the exact limits from CMS and ATLAS one has to perform a careful reconstruction and selection of events employing suitable cuts, and by taking into account the generator-level efficiency along with a background estimation, which is beyond the scope of this paper\footnote{A recent analysis can be found in~\cite{Belyaev:2020wok}.}. {Alternatively here we make an estimate of the lifetime of $H^\pm$ with the allowed values of $\Delta M$ and $M_{H^0}$
and project the available limits from CMS~\cite{CMS-PAS-EXO-19-010} to realize if at all it is feasible to see the charged tracks in colliders.}
This in turn could imply a collider probe for an alternative cosmological history of the Universe.  

As stated earlier, for $\Delta M\in\{140-200\}~\rm MeV$, $H^\pm$ decays dominantly into $\pi^\pm,H^0$ final state, while for $\Delta M<m_\pi$ the decay turns out to be semi-leptonic 3-body final state. In the top left panel of Fig.~\ref{fig:prodcs} we see a manifestation of this, where the branching $\text{Br}\left(H^\pm\to\pi^\pm,H^0\right)$ into pion final state decreases with the increase in $\Delta M$ as the 3-body decay starts dominating. Note that, in this case the DM mass also varies in the range $m_{H^0}\in\{450-463\}~\rm GeV$. Following Eq.~\eqref{eq:pi-life} we also expect, for large $\Delta M$, the lifetime $\tau_{H^\pm\to\pi^\pm H^0}$ should decrease producing a shorter disappearing track. This is exactly what we see in the top right panel of Fig.~\ref{fig:prodcs}. Thus, a larger $\text{Br}\left(H^\pm\to\pi^\pm,H^0\right)$ implies a longer lifetime $\tau_{H^\pm\to\pi^\pm H^0}$ (and a smaller $\Delta M$) or equivalently a longer track length. This, in turn, places constraints on the model parameter which we are going to discuss next. One should also note the presence of points satisfying relic abundance for $n=2$ with different choices of $T_R$ on the same plane, indicating the possibility of testing benchmark points obtained from the analysis in the last sections in collider experiments. 


In the bottom left panel of Fig.~\ref{fig:prodcs} we project the experimental limit~\cite{Sirunyan:2018ldc,CMS-PAS-EXO-19-010} from CMS on the decay lifetime of $H^\pm$ obtained using our model parameters. The red line corresponds to the CMS limit where the decaying charged particle has 100\% decay branching fraction into pion final state, whereas for the blue line the pion decay branching fraction is 95.5\%. The black thick curve shows the total lifetime of $H^\pm$ as a function of $m_{H^\pm}$  obtained numerically for a fixed DM mass of 450 GeV. We again show three benchmark points where observed relic density can be obtained for $n=2$ with different $T_R$. We note, based on the approximate analysis, $\Delta M\lesssim 200~\rm MeV$ is tightly constrained from CMS and likely to be ruled out, which also agrees with earlier observations~\cite{Belyaev:2016lok}. However, large $\Delta M(>200~\rm MeV)$ regions with shorter lifetime (for example the point denoted by ``X" in the bottom left panel of Fig.~\ref{fig:prodcs}) still can be seen lying beyond the reach of present CMS bound. It is understandable, by tuning $n,T_R$, it is always possible to accommodate points for $\Delta M>200$ MeV which satisfy relic density that are safe from CMS exclusion. 
We can thus infer, for any given $(n,T_R)$, the region of parameter space satisfying DM constraints with lifetime $\lesssim 0.1~\rm ns$ (equivalent to a track length of $\lesssim 1~\rm cm$) are beyond the present sensitivity of CMS experiment, and thus safe. Finally, in the bottom right panel we show the production cross-section for the processes $pp\to H^{+}H^{-},H^{\pm}H^0(A^0)$ at $\sqrt{s}=13~\rm TeV$. A detailed analysis utilizing the numerically obtained production cross section can constrain $m_{H^\pm}$ and therefore the DM mass, by providing the number of disappearing track events for a given luminosity. However, here we only show that our model parameters can give rise to a sizeable production cross-section in colliders abiding all DM constraints. For computing the production cross-section we have again relied upon {\tt CalcHEP}~\cite{Belyaev:2012qa} and used {\tt CTEQ6l} as the representative parton distribution function (PDF)~\cite{Placakyte:2011az}. We see, for DM mass $\gtrsim 400~\rm GeV$, the production cross-section is $\sim 2~\rm fb$. For all the plots, to show the corresponding DM parameter space, we have chosen $\lambda_L=0.01$ such that the DM is safe from direct and indirect search constraints. 
We conclude this section by observing that a charged track of length $\lesssim\mathcal{O}(1)~\rm cm$ could indeed be a probe for a non-standard cosmological parameters for the IDM providing an evidence for fast expanding pre-BBN era at the LHC.

}



\subsection{ITDM in a fast expanding Universe}\label{sec:itdm}

\begin{figure*}[htb!]
  \centering
  \subfigure[]{\includegraphics[scale=0.32]{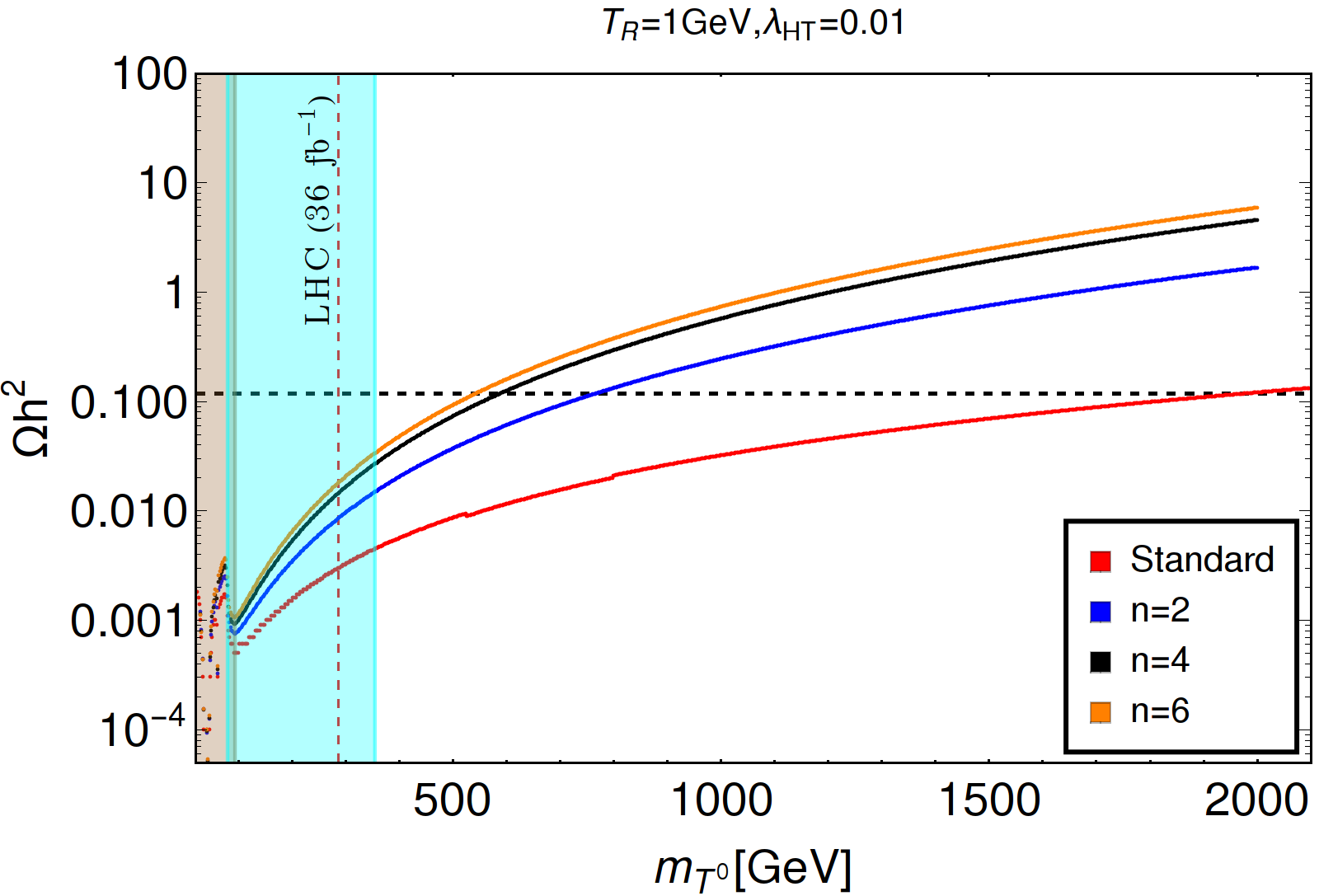}}\quad
  \subfigure[]{\includegraphics[scale=0.32]{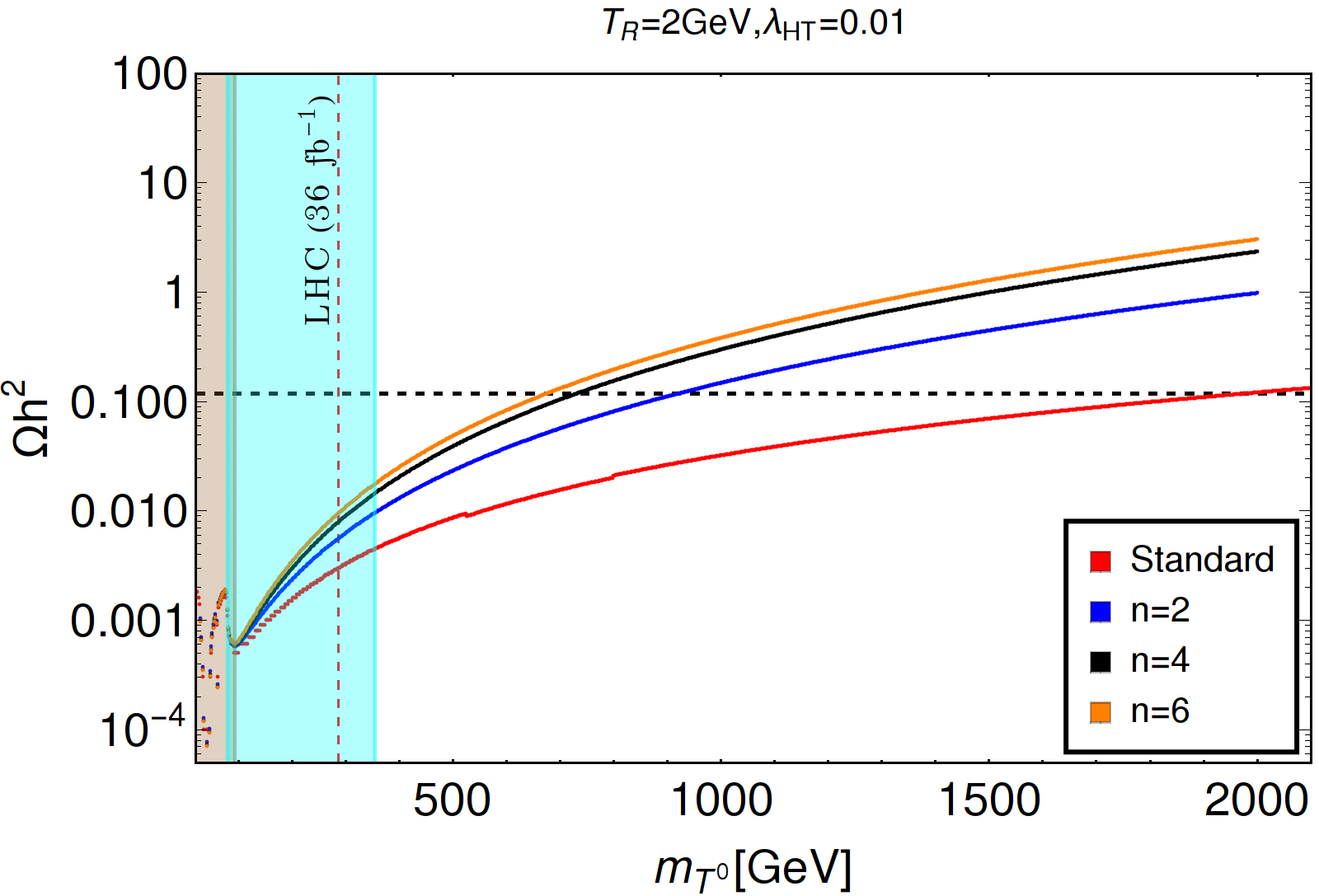}}\quad
 \subfigure[]{\includegraphics[scale=0.32]{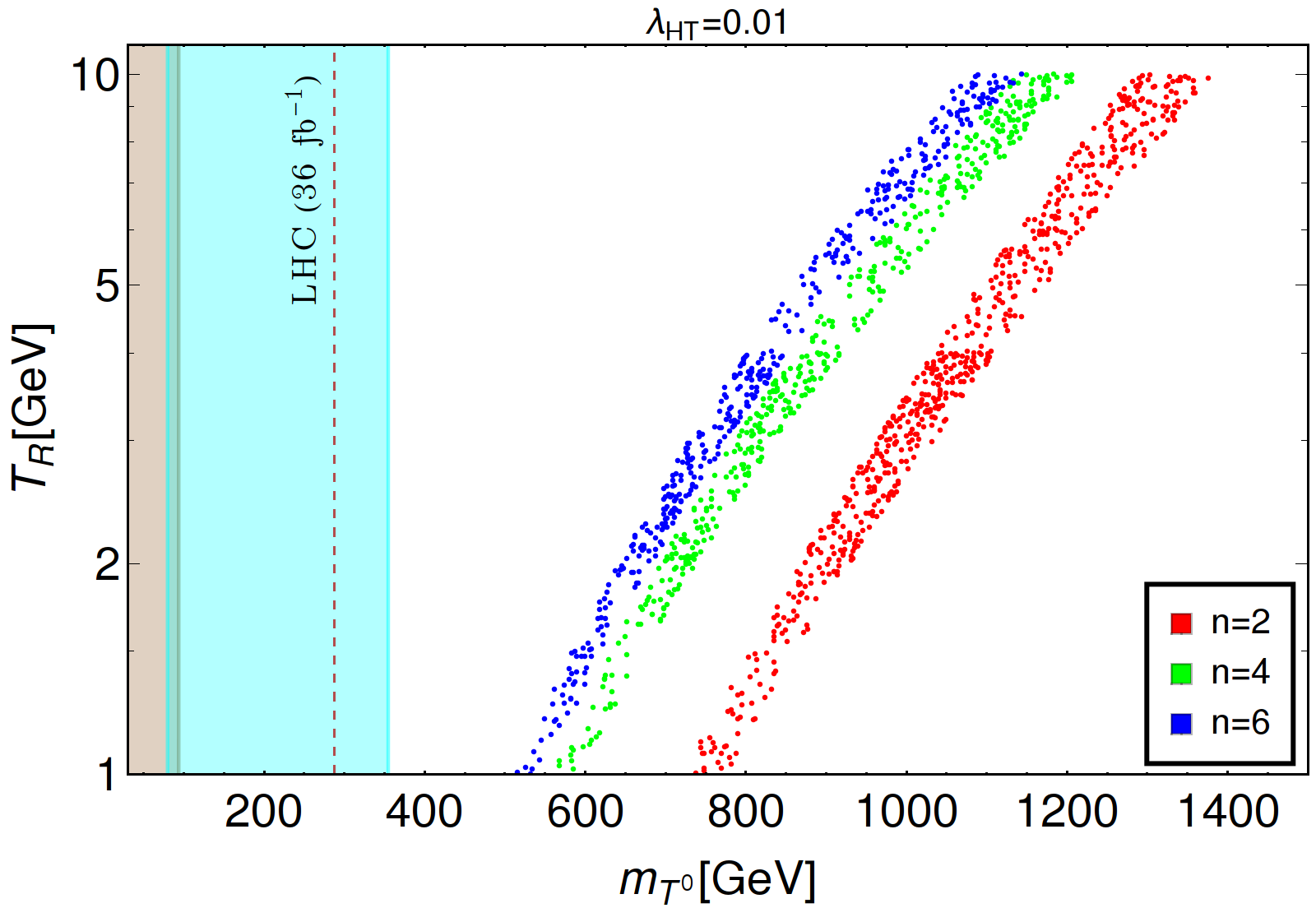}}\quad
  \caption{Top: Variation of DM relic abundance with ITDM mass where the colourful curves correspond to different values of $n$ for a fixed $T_R$ as mentioned in the plot legends. Bottom: Relic density allowed parameter space in $T_R-m_{T^0}$ plane for different choices of $n=2,4,6$ shown in respectively red, green and blue. The brown and the cyan regions respectively show the DM mass region disallowed by the direct search (XENON1T) and indirect search ($W^+W^-$ final state) data. The red dashed straight line in each plot shows the limit from LHC on triplet mass for $36~\text{fb}^{-1}$ of luminosity at $\sqrt{s}=13~\rm TeV$. In all cases we have set the portal coupling to a fixed value of $\lambda_{HT}=0.01$.}\label{fig:param-trip}
\end{figure*}

As mentioned in the beginning, in order to recover the {\it desert} region beyond the IDM paradigm, we also apply the prescription of modified Hubble rate due to fast expansion to scalar DM with larger representation under $SU(2)_L$. Here we describe the general structure of a $SU(2)_L$ triplet dark matter model. In this set-up the SM is extended by introducing a $SU(2)_L$ triplet scalar with hypercharge $Y=0$. An additional $Z_2$ symmetry is also imposed under which the triplet transforms non-trivially. It is also considered that 
the triplet has zero VEV. The scalar potential under $\text{SM}\times Z_2$ symmetry then reads~\cite{Araki:2011hm}  

\bea\begin{aligned}
& V\left(H,T\right)\supset\mu_H^2 \left|H\right|^2+\lambda_H\left|H\right|^4+ \frac{\mu_T^2}{2}\text{Tr}\Bigl[T^2\Bigr]\\&+\frac{\lambda_T}{4!}\Biggl(\text{Tr}\Bigl[T^2\Bigr]\Biggr)^2+\frac{\lambda_{HT}}{2}\left|H\right|^2\text{Tr}\left[T^2\right],
    \end{aligned}\label{eq:pot-trip}
\eea

\noindent where $H$ is the SM-like Higgs doublet and the triplet $T$ is parameterized as 

\bea
T=\begin{pmatrix}
T^0/\sqrt{2}&&-T^+ \\ -T^-&&-T^0/\sqrt{2}
\end{pmatrix}.
\eea

\noindent Now, after electroweak symmetry breaking the masses of the physical scalar triplets are given by

\bea
m_{T^0,T^\pm}^2 = \mu_T^2+\frac{\lambda_{HT}}{2}v^2,\label{eq:trip-mass}
\eea

\noindent with $v=246~\rm GeV$. Notice that although mass of neutral and charged triplet scalar are degenerate (Eq.~\eqref{eq:trip-mass}), a small mass difference $\delta m\simeq 166~\rm MeV$ can be generated via 1-loop radiative correction~\cite{Cirelli:2009uv} that makes $T^0$ as the lighter component and hence a stable DM candidate. This is the crucial difference between IDM and scalar triplet DM where in IDM the mass difference is a free parameter while for scalar triplet this is fixed from 1-loop correction. The bounded from below conditions for the scalar potential in all field directions in Eq.~\eqref{eq:pot-trip} require

\bea\begin{aligned}
& \lambda_{H,T}\geq 0;~~ \sqrt{\lambda_H\lambda_T}>\frac{1}{2}\left|\lambda_{HT}\right|.    
    \end{aligned}
\eea

Apart from the theoretical constraints arising from the stability, perturbativity and tree-level unitarity of the scalar potential one needs to also consider the experimental constraints on the parameters of the scalar potential. As the charged and neutral component of the triplet scalar are almost degenerate, the contributions to the $T$ and $U$ parameters are very much suppressed in this scenario. However, the charged component $T^\pm$ can contribute significantly to the Higgs diphoton signal strength which is accurately measured $\mu_{\gamma\gamma}=0.99\pm 0.14$ from ATLAS~\cite{Aaboud:2018xdt} and $\mu_{\gamma\gamma}=1.17\pm 0.10$ from CMS. It has recently been shown~\cite{Chiang:2020rcv,Bell:2020hnr} that searches for disappearing tracks at the LHC excludes a real triplet scalar lighter than 287 GeV using $36~\text{fb}^{-1}$ of data at $\sqrt{s}=13~\rm TeV$.

\begin{figure*}[htb!]
$$
  \includegraphics[scale=0.45]{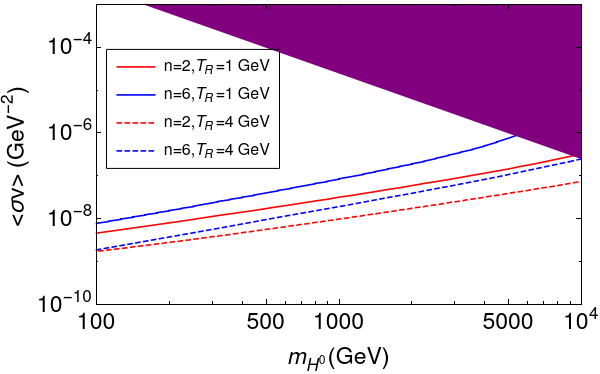}
 $$
  \caption{Required order of cross-sections to satisfy the observed DM abundance for different choices of $\{n,T_R\}$ are shown as function of DM mass. The purple region is disfavored by the perturbative unitarity bound on DM pair annihilation cross-section (see text for details).}\label{fig:per-bound}
\end{figure*}
 
We again numerically solve the BEQ in Eq.~\eqref{eq:BoltzDM2} with the modified Hubble rate in Eq.~\eqref{eq:mod-hubl} and determine the subsequent DM relic density for different choices of the fast expansion parameters $n,T_R$. In the top and middle panel of Fig.~\ref{fig:param-trip} we show the variation of the DM relic abundance as a function of the ITDM mass. Here we have kept the portal coupling fixed and obtained the resulting direct and indirect search exclusion regions for $\lambda_{HT}=0.01$. The parameter space excluded by XENON1T limit is shown by the brown region where the direct search cross-section is given by~\cite{DuttaBanik:2020jrj}

\bea
\sigma_{n-T^0}^\text{SI} = \frac{\lambda_{HT}^2 f_N^2}{4\pi}\frac{\mu^2 m_n^2}{m_h^4 m_{T^0}^2}
\eea

\noindent while the indirect search exclusion due to the $W^+W^-$ final state is shown by the cyan region. Since the mass splitting $\delta m$ is no more a free parameter and fixed to a small value of $\delta m\simeq 166~\rm MeV$, co-annihilation plays the dominating role here. As a result right relic is obtained in the case of ITDM for a very large DM mass $m_{T^0}\sim 1.8~\rm TeV$ as shown by the red curve $(n=0)$ in each plot. Once fast expansion is introduced, there is drastic improvement in the parameter space. As one can see, for $T_R=1~\rm GeV$ right relic density is achievable for $m_{T^0}\sim 800~\rm GeV$ with $n=2$ (blue curve). While for $T_R=2$ GeV, the relic satisfied mass is around 900 GeV with $n=2$. As inferred earlier, this happens because for smaller $T_R$ the expansion rate increases following Eq.~\eqref{eq:mod-hubl}. This is being compensated by a smaller choice of the DM mass to satisfy the observed abundance since $\langle\sigma v\rangle\propto 1/m_{T^0}^{2}$. Enhancement of $n$ could provide further smaller relic satisfied DM mass consistent with direct, indirect and LHC searches.
Varying $\lambda_{HT}$ would give similar results since the effective annihilation cross section is mostly dominated by gauged mediated co-annihilation hence almost insensitive to $\lambda_{HT}$ unless it is very large ($\gtrsim 0.1$) which is anyway disfavored from direct and indirect search bounds. In the bottom panel of Fig.~\ref{fig:param-trip} we vary the DM mass $m_{T^0}$ by keeping $\lambda_{HT}=0.01$ and obtain the resulting relic abundance allowed parameter space in $T_R-m_{T^0}$ plane for different choices of $n$. Here we again see the manifestation of faster expansion elaborated above {\it i.e.,} for a fixed DM mass, a smaller $n$ (in red) needs a smaller $T_R$ in order to obtain the observed relic density.  Note that in all cases we have considered $T_R\geq 1~\rm GeV$ to ensure that ITDM remains in thermal equilibrium at high temperature. Limits from direct, indirect and LHC searches are also projected with the same colour code as before. Taking all relevant constraints into account, we see from bottom panel of Fig.~\ref{fig:param-trip}, the region $m_{T^0}\gtrsim 450$ GeV can be recovered considering $2\leq n \leq 6$ and $T_R\gtrsim$ 1 GeV. We find, it is also possible to resurrect part of the parameter space below 450 GeV for $T_R<1~\rm GeV$ ensuring the DM thermalizes in the early Universe depending on the choice of $n$. This is, however, in contrast to the case of IDM dark matter, where the lower bound on the allowed DM mass ($\gtrsim 350~\rm GeV$), satisfying thermalization criteria, is almost independent of the fast expanding parameters.


The discovery prospects for a real triplet extension of the SM at the colliders have been discussed in~\cite{Chiang:2020rcv,Bell:2020hnr}. As inferred in~\cite{Chiang:2020rcv}, the $13~\rm TeV$ LHC excludes a real triplet lighter than $\{287,608,761\}~\rm GeV$ for $\mathcal{L}=\{36,300,3000\}~\text{fb}^{-1}$ of luminosity. The present case where the neutral triplet scalar is stable and contributes to the DM Universe (ITDM) can be probed at the colliders via disappearing track signature through the decay of the long-lived charged component: $T^\pm\to\pi^\pm T^0$ due to small mass splitting $\delta m$. The situation is exactly similar as that in the case of IDM dark matter discussed in Sec.~\ref{sec:collider}, hence we do not further repeat it here. 

The requirement of perturbative unitarity of the DM annihilation cross-section can forbid some part of the relic density allowed parameter space depending on the choice of  $\{n,T_R\}$~\cite{DEramo:2017gpl}, thus providing a bound on the DM mass. A general prescription for obtaining upper bound on thermal dark matter mass  using such partial wave unitarity analysis has been worked out in~\cite{Griest:1989wd}. The upper limit on the thermally averaged DM interaction cross section is provided by~\cite{Griest:1989wd}

\begin{equation}
\langle\sigma v\rangle_\text{max} \lesssim \frac{4\pi}{m_{\rm DM}^2}\sqrt{\frac{x_f}{\pi}}  
\end{equation}

\noindent By using the approximate analytical estimate of DM yield in  Eq.(\ref{eq:mod-yld}) (following the approach in \cite{Kolb:1990vq}), the freeze-out temperature $x_f$ can be approximately determined by using the semi-analytical expression for DM yield by equating DM abundances before and after freeze out (see Eq.(\ref{eq:abunA}))

\begin{align}
    e^{x_f} x_f^{1/2}\simeq \frac{c(c+2)}{c+1}\times \frac{0.192 ~M_{\rm pl}}{g_*^{1/2}}\frac{\langle\sigma v\rangle m_{\rm DM}}{\left(\frac{x_r}{x}\right)^{n/2}},
\end{align}

\noindent with $c\sim\mathcal{O}(1)$ constant. We calculate the annihilation cross-section that gives rise to right relic abundance numerically, and compare that with the maximum cross-section allowed by the partial wave unitarity. This eliminates a part of the parameter space for a fixed DM interaction rate, as shown by the purple region in Fig.~\ref{fig:per-bound}. It turns out that for both IDM~\cite{LopezHonorez:2010tb} and ITDM~\cite{Ayazi:2014tha}, the leading contribution to the DM annihilation cross-section is $s$-wave dominated. We find, regions with large $n\geq 2$ (and small $T_R$) are typically in tension with the unitarity bound at the higher range of DM mass. This is expected, since for large $n$ (or small $T_R)$ the Hubble parameter is large, hence the interaction rate needs to be larger to avoid over abundance. This is in conflict with the maximum allowed annihilation cross-section, disfavouring large $\langle\sigma v\rangle$. Now, for the case of IDM, we are specifically interested in the mass window $m_W\lesssim m_{H^0}\lesssim 525~\rm GeV$ while for ITDM $m_{T^0}\lesssim 2~\rm TeV$. On the other hand, as explained earlier, we choose $n\leq 6$ ($T_R\gtrsim 1$ GeV) to ensure that the DM thermalizes in the early Universe above the weak scale. Thus, within our working regime of $n$ and $T_R$, we find the partial wave unitarity bound does not pose any serious constraint for the DM mass range of our interest.

\section{Conclusion}\label{sec:concl}

In this work, considering a form of alternative cosmology, we revisit two popular DM scenarios where the DM is part of $SU(2)_L$ multiplets (other than singlet). We first take up the minimal inert doublet model (IDM) where it is observed that an intermediate DM mass range: $80~\text{GeV}\lesssim m_\text{DM}\lesssim 525~\rm GeV$ is disfavored in a radiation dominated Universe due to relic under abundance via freeze-out. In an attempt to circumvent this, extension of the minimal inert doublet model or existence of multiple DM candidates have been proposed earlier. Here, we follow a different route and find that without resorting to an extended particle spectrum revival of the {\it desert region} is possible in presence of a non standard epoch in the early Universe. We obtain the parameter space accounting for the correct relic abundance for single component inert doublet DM by varying the relevant parameters responsible for the fast expansion of the Universe. Subsequently, we see that a major part of the relic density allowed region gets ruled out from DM direct and indirect search constraints and this in turn puts a restriction on the fast expansion parameters. In particular, we found that for $\lambda_L=0.01$, the DM mass below $350$~GeV is ruled out irrespective of the cosmological history of the early Universe. The bound turns severe for larger $\lambda_L$ {\it i.e.,} for higher interaction rate.  While for pure IDM, bounds from relic density and (in)direct search experiments do not allow a large mass splitting, for inert scalar triplet, on the other hand, this happens naturally due to small radiative mass splitting. We then discuss possible collider signature for pure IDM under the influence of fast expansion, and find that the newly obtained parameter space can be probed via the identification of the charged track signal of a long-lived charged scalar. The resulting track length depends on the mass splitting between the charged and neutral component of the inert scalar doublet. The track length  $(\lesssim\mathcal{O}\left(1\right)~\rm cm)$ for such a long-lived scalar, however, is below the sensitivity from the present CMS/ATLAS search and hence leaves the possibility of being probed in future experiments. This also implies the prospect of probing the modified cosmological history of the Universe in collider experiments. 

We extend our analysis by applying the same methodology to scrutinize the case for hyper-chargeless real triplet scalar DM anticipating such modification in DM parameter space should also be observed for larger representation of the DM field. We show that a significant parameter space ($m_{T^0} \gtrsim 450$ GeV considering $2\leq n\leq 6$ and $T_R\gtrsim 1$ GeV) satisfying the relic density and other DM search bounds for $m_{T^0}\lesssim 2$ TeV and portal coupling $\lambda_{HT}=0.01$ can indeed be restored for the scalar triplet scenario, which is otherwise disallowed. We thus conclude, this prescription can be applied for any DM candidate which is a part of a $SU(N)$ multiplet or even for different multi component DM frameworks. Implications of our analysis on different aspects of particle physics and cosmology such as electroweak phase transitions, prediction of gravitational waves, neutrino physics and leptogenesis remain open. We keep these studies for future endeavours. 
\section*{Acknowledgement}
One of the authors, AKS appreciates Sudipta Show for several discussions during the course of work. AKS is supported by NPDF grant PDF/2020/000797 from Science and Engineering Research Board, Government of India. PG would like to acknowledge the support from DAE, India for the Regional Centre for Accelerator based Particle Physics (RECAPP), Harish Chandra Research Institute. FSQ is supported by the Sao Paulo Research Foundation (FAPESP) through grant 2015/15897-1 and ICTP-SAIFR FAPESP grant 2016/01343-7. FSQ acknowledges support from CNPq grants 303817/2018-6
and 421952/2018-0 and the Serrapilheira Institute (grant
number Serra-1912-31613).

\appendix
\section{Semi-analytical freeze-out yield}\label{sec:analyt-yld}

To obtain a semi-analytical expression for the DM yield under the influence of fast expansion we closely follow Ref.~\cite{DEramo:2017gpl}. Assuming the DM freezes out during the epoch of $\eta$-domination {\it i.e.,} $x_f\ll x_r$ the BEQ in Eq.~\eqref{eq:BoltzDM2} can be approximated as

\bea\begin{aligned}
&  \frac{dY_\text{DM}}{dx}\simeq -A\frac{\langle\sigma v\rangle }{x^{2-n/2}x_r^{n/2}}\Bigl(Y_\text{DM}^2-Y_{\rm DM}^{\rm eq^2}\Bigr)
\end{aligned}
\eea

\noindent with $A=\frac{2\sqrt{2}\pi}{3\sqrt{5}}\sqrt{g_*}m_\text{DM}M_\text{pl}$. Defining $\Delta\equiv Y_\text{DM}-Y_\text{DM}^\text{eq}$ and ignoring terms proportional to $\mathcal{O}\left[\Delta^2\right]$ in times much earlier than freeze-out (as departure from equilibrium is minimal), while neglecting the equilibrium distribution in the post freeze-out regime, we obtain

\begin{align}
&\ Y_\text{DM}\left(x\right)\simeq
\begin{cases}
      Y_\text{DM}^\text{eq}\left(x\right)+\frac{x^{2-n/2}x_r^{n/2}}{2A\langle\sigma v\rangle} \text{~~~~for~~} ~1<x<x_f\\
      \Biggl(\frac{1}{Y_\text{DM}\left(x_f\right)}+A\,\xi\left(x\right)\Biggr)^{-1}\text{~for~~} ~x_f<x<x_r 
    \end{cases} \label{eq:abunA}
\end{align}

where

\bea
\xi\left(x\right) = \frac{1}{x_r^{n/2}}\int_{x_f}^x\,dx\,\frac{\langle\sigma v\rangle}{x^{2-n/2}}.\label{eq:xi-int} 
\eea

Now, one can expand the thermally averaged cross-section in terms of the partial waves as: $\langle\sigma v\rangle\simeq\sigma_s+\sigma_p/x+\mathcal{O}\left(x^{-2}\right)$. Considering $s$-wave domination and on substitution in Eq.~\eqref{eq:xi-int} we find 
\begin{align}
&\xi\left(x\right) = \frac{\sigma_s}{x_r^{n/2}}
\begin{cases}
      \frac{x_f^{n/2-1}-x^{n/2-1}}{1-n/2} \text{~~~~for~~}n\neq 2\\
    ~ \text{Log}\left[\frac{x}{x_f}\right] \text{~~~~~~~~for~~}n= 2     \end{cases}  
\end{align}
After the end of fast expansion regime ($x>x_r$), the radiation dominates the energy density and the resulting DM yield reads

\bea\begin{aligned}
& Y_\text{DM}\left(x\right)\simeq \Biggl(\frac{1}{Y_\text{DM}\left(x_f\right)}+A\,\xi_\text{rad}\left(x\right)\Biggr)^{-1},~x>x_r     
\end{aligned}
\eea

\noindent where

\bea
\xi_\text{rad}\left(x\right) = \int_{x_r}^x\,dx\,\frac{\langle\sigma v\rangle}{x^2}.
\eea




\section{BBN constraints}\label{sec:bbn}
The effect of the new species $\eta$ can be parametrized by an effective number of relativistic degrees of freedom (DOF) as evident from Eq.(\ref{eq:totED}).

\bea\begin{aligned}
& \rho\left(T\right) = \frac{\pi^2}{30}g_{*\text{eff}} T^4     \end{aligned}\eea

\noindent with 

\bea\begin{aligned}
& g_{*\text{eff}} = g_*^\text{SM}+\Delta g_*^\eta \\&=\Bigl(2+\frac{7}{8}\times 4\Bigr)+\Bigl(2\times\frac{7}{8}\times N_\nu\Bigr)+\Bigl(2\times\frac{7}{8}\times \Delta N_\nu\Bigr),  
\end{aligned}
\eea

\noindent The first two terms in the last equation stand for the SM contribution with the $N_\nu$ indicates the number of effective neutrinos. The notation $\Delta N_\nu$ accounts for the $\eta$ contribution to the number of relativistic degrees of freedom as obtained from Eq.(\ref{eq:totED}).

\bea\begin{aligned}
& \Delta N_\nu  = \frac{4}{7}\, g_*\left(T_R\right)\,\Biggl(\frac{g_{* s}\left(T\right)}{g_{* s}\left(T_R\right)}\Biggr)^{(4+n)/3}\,\Biggl(\frac{T}{T_R}\Biggr)^n.   
\end{aligned}
\eea

\noindent Considering $T_R$ around $T_\text{BBN}$ and $T\sim T_{\rm BBN}$ we can assume $g_{*s}(T)\sim g_{*s}(T_R)$. We also use $g_{*}(T_R)=\left(2+\frac{7}{8}\times 4+\frac{7}{8}\times 2\times 3\right)$ to include the contributions of photon, positrons and neutrinos and reach at

\bea\begin{aligned}
& \Delta N_\nu\simeq 6.14\Biggl (\frac{T}{T_R}\Biggr)^n.    
\end{aligned}
\eea

\noindent Since the additional contribution to $N_\nu$ is positive, we use the bound $N_\nu+\Delta N_\nu\lesssim 3.4$~\cite{Cyburt:2015mya} at 95\% CL (2$\sigma$) and $T\simeq 1~\rm MeV$ to obtain

\bea
T_R \gtrsim \left(15.4\right)^{1/n}~\text{MeV}.
\eea

\bibliographystyle{JHEP}
\bibliography{Bibliography}
\end{document}